\documentclass[10pt,preprint]{aastex}
\usepackage{natbib,apjfonts,psfig,rotating,graphicx}

\newcommand{\etal} {et~al.}
\def\spose#1{\hbox to 0pt{#1\hss}}
\newcommand\lsim{\mathrel{\spose{\lower 3pt\hbox{$\mathchar"218$}}
     \raise 2.0pt\hbox{$\mathchar"13C$}}}
\newcommand\gsim{\mathrel{\spose{\lower 3pt\hbox{$\mathchar"218$}}
     \raise 2.0pt\hbox{$\mathchar"13E$}}}
\newcommand{\agn}{{\small AGN}}

\newcommand{\chandra}{{\it Chandra}}

\newcommand{\emm}{{\small EM}}

\newcommand{\fac}{{\small FAC}}

\newcommand{\gsrp}{{\small GSRP}}

\newcommand{\nasa}{{\small NASA}}

\newcommand{\ngc}{{\small NGC}}
\newcommand{\nsf}{{\small NSF}}

\newcommand{\photoion}{{\small PHOTOION}}

\newcommand{\rrc}{{\small RRC}}

\newcommand{\xmm}{{\it XMM-Newton}}
\newcommand{\xspec}{{\small XSPEC}}

\begin{document}

\title{Atomic Calculations and Spectral Models of X-ray Absorption and Emission Features From Astrophysical Photoionized Plasmas}

\author{A. Kinkhabwala\altaffilmark{1}, E. Behar\altaffilmark{1,2}, M.
Sako\altaffilmark{1,3,4}, M.F. Gu\altaffilmark{4,5}, 
S.M. Kahn\altaffilmark{1}, F.B.S. Paerels\altaffilmark{1}}

\altaffiltext{1}{Columbia Astrophysics Laboratory, 
	      Columbia University, 
	      538 West 120th Street, 
	      New York, NY 10027; 
	      ali@astro.columbia.edu,
              behar@astro.columbia.edu,
              masao@astro.columbia.edu,
              skahn@astro.columbia.edu,
              frits@astro.columbia.edu}
\altaffiltext{2}{Present Address: Physics Department, 
                Technion, Haifa 32000, 
                Israel;
                behar@physics.technion.ac.il}
\altaffiltext{3}{Present Address: Theoretical Astrophysics and Space 
                Radiation Laboratory,
		California Institute of Technology,
		MC 130-33,
		Pasadena, CA 91125;
		masao@tapir.caltech.edu}
\altaffiltext{4}{Chandra fellow}
\altaffiltext{5}{Center for Space Research,
	      	Massachusetts Institute of Technology,
		Cambridge, MA, 02139;
		mfgu@space.mit.edu}

\shorttitle{Photoionized Plasmas}
\shortauthors{Kinkhabwala \etal}

\received{}
\revised{}
\accepted{}


\begin{abstract}

We present a detailed model of the discrete X-ray spectroscopic features 
expected from steady-state, low-density photoionized plasmas.  
We apply the Flexible Atomic Code (FAC) to calculate all of the
necessary atomic data for the full range of ions relevant for the X-ray 
regime.  These calculations have been incorporated into a 
simple model of a cone of ions irradiated by a point source located at its 
tip (now available as the \xspec\ model \photoion).  For each ionic species 
in the cone, photoionization is balanced by recombination and ensuing 
radiative cascades, and photoexcitation 
of resonance transitions is balanced by radiative decay.  This simple model is 
useful for diagnosing X-ray emission mechanisms, determining 
photoionization/photoexcitation/recombination rates, fitting temperatures
and ionic emission measures, and probing geometrical properties 
(covering factor/column densities/radial filling factor/velocity 
distributions) of absorbing/reemitting regions in photoionized plasmas.
Such plasmas have already been observed in diverse astrophysical X-ray 
sources, including active galactic nuclei, X-ray binaries, cataclysmic 
variables, and stellar winds of early-type stars, and may also provide a 
significant contribution to the X-ray spectra of gamma-ray-burst afterglows 
and the intergalactic medium.

\end{abstract}


 

\section{Introduction}\label{sec:intro}

With the launch of the X-ray satellites \chandra\ (Weisskopf \etal\ 2002) and 
\xmm\ (Jansen \etal\ 2001; den Herder \etal\ 2001), high-resolution X-ray 
spectroscopy of diverse astrophysical objects has now become routine.
X-rays, due to their origin in extreme astrophysical environments as 
well as their significant penetrating ability, can potentially provide a 
wealth of information about these sources.   Their penetration power, in 
particular, implies that X-ray emitting plasmas are often optically thin.  In 
such cases, complicated radiative transfer is negligible, allowing for 
particularly simple astrophysical models.

There are two main types of X-ray line-emitting plasmas:
(1) Hot plasmas, which are mechanically heated through collisions and
therefore have temperatures comparable to the observed line energies, and which
produce X-ray line emission predominantly through radiative decay following 
electron impact excitation (hereafter, ``collision-driven''), and
(2) Photoionized plasmas, which are irradiated by a powerful external 
source and have subsequently lower temperatures consistent with 
photon heating (including both photoelectric and Compton heating),
and which produce X-ray line emission through recombination/radiative cascade 
following photoionization, and radiative decay following photoexcitation 
(hereafter, ``radiation-driven''\footnote{``Radiation-driven'' often refers 
to the force due to radiation pressure, but throughout this paper, this term 
is used only to refer to the ultimate power source for driving atomic 
transitions, subsuming both photoionization and photoexcitation.} or 
``photoionized'').
Hot plasmas are produced in the coronae of stars (including the Sun), in 
shock-heated environments (such as supernovae, cataclysmic variables, and 
stellar winds of early-type stars), and in the intracluster media of clusters 
of galaxies.  
Photoionized plasmas have recently been unambiguously confirmed in X-ray 
binaries (Liedahl \& Paerels 1996, Cottam \etal\ 2001), \agn\ outflows 
(Sako \etal\ 2000, Kinkhabwala \etal\ 2002), cataclysmic variables 
(Mukai \etal\ 2003), and even in the stellar wind of the WR+O binary $\gamma$ 
Velorum (Dumm \etal\ 2002).  Hybrid plasmas, in which 
both collision-driven and radiation-driven processes are important, can also
exist.  
So far, hybrid X-ray-line-emitting plasmas have not yet been unambiguously 
detected in any astrophysical source, but they are expected to play an 
important role in accretion disks, for example, which are strongly irradiated 
(implying radiation-driven processes are important) but also have high 
densities (implying collision-driven processes are important as well).  
Finally, we note that the mechanisms driving X-ray emission (either 
collision-driven or radiation-driven) in the two vastly different regimes of 
gamma-ray-burst (GRB) afterglows and the intergalactic medium (IGM) are still
largely unknown.  GRB afterglows may harbor hot plasmas produced through 
shocks and/or photoionized plasmas located in the ejecta or
surrounding interstellar medium and powered by radiation from the 
burst and afterglow radiation.   Similarly, the IGM may harbor hot plasmas 
from shocks due to structure formation and/or photoionized plasmas powered by 
radiation from active galactic nuclei (AGN) or starbursts.

Calculations of X-ray line emission from hot, collision-driven plasmas 
have been discussed in great detail by several authors, including Raymond \& 
Smith (1977), Mewe, Gronenschild, \& van den Ooord (1985), Mewe, Lemen, \& van 
den Oord (1986), Kaastra (1992), Liedahl, Osterheld, \& Goldstein (1995), and 
Smith \etal\ (2001).  Discussions of line formation in X-ray photoionized 
plasmas have also been presented by several authors, including
Tarter, Tucker, \& Salpeter 
(1969), Halpern \& Grindlay (1980), Krolik, McKee, \& Tarter (1981), 
Kallman \& McCray (1982), Netzer (1993), Kallman \etal\ (1996), and Liedahl 
(1999); however, 
spectral predictions made in these works have not yet been compared in detail
with real astrophysical X-ray spectra.  In contrast, the 
assumptions and calculations we describe here for photoionized plasmas 
were made and tested in the 
process of analyzing high-resolution X-ray spectra from various astrophysical 
sources.

Below, we present atomic structure and transition calculations using the 
publicly-available atomic code \fac\ (Gu 2002; Gu 2003) for all ionic 
transitions relevant for the X-ray regime.  Atomic data values for some
especially important 
transitions (in particular, more accurate wavelengths) and for some 
photoelectric edges are taken from other atomic databases, as will be described
below.  Using these 
data, we have fully characterized all significant line and edge absorption 
from the full range of included ions.  We have also fully 
characterized the X-ray spectral emission features for radiative recombination 
forming the important H-like and He-like ions, as well as both radiative and 
dielectronic recombination forming L-shell ions.

In order to attempt to explain real astrophysical spectra from photoionized  
plasmas, we have created a simple model of a cone of ions irradiated by a 
source located at its tip.  Photoionization in the cone is balanced by 
recombination/radiative cascade and photoexcitation is balanced by radiative 
decay.  This model has been incorporated into \xspec\ (Arnaud 1996) as the 
additive model 
\photoion\footnote{Available at http://xmm.astro.columbia.edu/research.html}.  
For convenience, we have also created the abbreviated additive model 
{\small PHSI} for a single ion, and the multiplicative models {\small MPABS} 
(multi-ion absorption) and {\small SIABS} (single-ion absorption) for pure 
absorption studies.

Our calculations invoke the following basic assumptions: 
(1)  All ions are in their ground states.  X-ray transitions are generally 
much faster than the relevant rates for radiation-driven and collision-driven 
processes, making this typically a safe assumption.  
(2)  Collisional excitation and collisional ionization are negligible.  
Due to the low temperatures and densities of photoionized plasmas, 
these processes are typically insignificant for driving X-ray transitions.
(3)  We assume the entire medium is in a global steady state.
(Note that we do not require the more restrictive assumption of {\em local} 
steady state conditions everywhere in the medium.)  Furthermore, for 
the medium as a whole, we assume the total ionization rate equals the 
total recombination rate between all {\em neighboring} charge states.
(4)  All relative ion ratios are constant throughout the medium.  Or, more
approximately, all ion ratios are constant over regions with size comparable to
the continuum optical-depth length scale (defined as the length over which the 
continuum is reduced by a few or more percent).  For a mildly-absorbed medium,
this length scale is equal to the entire medium, and this assumption is
trivially valid.  Interestingly, recent observations of the prototypical 
Seyfert~2 galaxy, \ngc~1068, show 
that an inherent, relatively scale-free density 
distribution (over a few orders of magnitude) at each radius is more important 
than the radial distribution in setting the range in ionization parameter 
(Brinkman \etal\ 2002; Ogle \etal\ 2003), suggesting that even in 
moderately or highly absorbed media this approximation may still be valid.
(5)  The irradiated medium can be modeled as a relatively narrow isotropic 
cone with the source of radiation located at its tip.  This cone is completely 
specified by its opening angle, ionic column densities, representative ionic 
recombination temperatures, and a simple velocity structure, characterized by 
a global radial velocity shift and gaussian width, and a global transverse 
(perpendicular to the cone) velocity shift and gaussian width.
(6)  The emitting plasmas are optically thin to their own emission.  
Because X-ray radiative cross-sections are generally weaker than longer 
wavelength cross-sections (e.g., in the UV or optical), a given medium remains 
optically thin to X-rays out to significantly higher column densities.  
Furthermore, in the context of our assumption of a {\em narrow} cone geometry, 
even with large radial ionic column densities and corresponding radial optical 
depth, the transverse optical depth is smaller than the radial optical depth 
by roughly the length-to-width ratio of the cone.  Observationally, optical 
thinness has been demonstrated for the line emitting regions of 
\ngc~1068 using a simple line conversion 
diagnostic (see Kinkhabwala \etal\ 2002 and \S\ref{sec:reemission} below).

Our paper is organized as follows.  Starting from the most general 
ionic rate equations in \S\ref{sec:master}, we describe how the 
above assumptions are employed to simplify these equations for generic 
astrophysical 
plasmas in \S\ref{sec:simp} and then for purely radiation-driven plasmas in 
\S\ref{sec:rad}.  Here, we also describe our calculations of important 
diagnostics for H- and He-like ion emission, as well as the specifics of all 
of our atomic calculations with FAC.  In order to explain real astrophysical 
spectra, we present our model of a cone of ions irradiated by a point source 
located at its tip in \S\ref{sec:cone}, obtaining the final expression for the 
rate equations, which serve as the basis for \photoion.  In 
\S\ref{sec:examples}, we provide examples of the capabilities of 
\photoion.  Finally, in \S\ref{sec:dis}, we discuss our results.

\section{Reaction Kinetics}\label{sec:master}

Presented below is a completely general analysis of the ionic rate equations for 
astrophysical plasmas (including all possible collision-driven and 
radiation-driven processes).  This analysis is undertaken for two reasons.  
First, we are 
unaware of any similarly general analysis in the literature and hope to 
motivate studies along similar lines.  
And, second, this 
analysis most clearly brings out all the assumptions that are made in the 
formulation of our final rate equations for a radiation-driven medium.

An extremely general form for the rate equation pertaining to 
population/depletion of ions $n_{Z,z,i}$ of element $Z$ (atomic number) 
with $z$ electrons in atomic level $i$ and immersed in a generic gas or 
plasma is:
\begin{eqnarray}
\lefteqn{\frac{dn_{Z,z,i}}{dt} = \sum_{k\geq1}\sum_j n_{z-1,j} n_{\mathrm{e}}^k C_{z-k,j}^{z,i}
+\sum_{1\leq k\leq z} \sum_j n_{z-k,j} \big[\{nC\}_Q\big]_{z-k,j}^{z,i}-\sum_{1\leq k\leq z}\sum_j n_{z,i} \big[R+D+n_{\mathrm{e}} C+\{n C\}_Q\big]_{z,i}^{z-k,j}} \nonumber\\
&\displaystyle{-n_{z,i} \sum_j \big[R+n_{\mathrm{e}} C+\{n C\}_Q+A\big]_{z,i}^{z,j} +\sum_j
n_{z,j}\big[R+n_{\mathrm{e}} C+\{n C\}_Q+A\big]_{z,j}^{z,i}}\nonumber\\
&\displaystyle{-n_{z,i}\sum_{k\geq1}\sum_j n_{\mathrm{e}}^kC_{z,i}^{z+k,j}-n_{z,i}\sum_{k\geq 1} \sum_j \big[\{nC\}_Q\big]^{z+k,j}_{z,i}+ \sum_{k\geq1}\sum_{j} n_{z+k,j}\big[R+D+n_{\mathrm{e}} C+\{n
C\}_Q\big]_{z+k,j}^{z,i}}\nonumber\\
&\displaystyle{+\sum_{q,q'}n_q\big[R+n_{\mathrm{e}}C+\{nC\}_Q\big]_{q}^{q'+z,i}-\sum_{q,q'}n_{z,i}\big[\{nC\}_q\big]_{q+z,i}^{q'}+\sum_{q,q'}n_qW_{q}^{q'+z,i}.}
\label{eq:master}
\end{eqnarray}
Because only one element at a time is concerned, we have dropped the 
atomic number $Z$ in the ion density subscripts on the RHS of the equation
for conciseness, i.e., $n_{z,j}\equiv n_{Z,z,j}$.  The full time derivative on 
the LHS can be written as the usual convective derivative:
$dn/dt={\partial n/\partial t}+{\mathrm{\bf v}}\cdot\nabla n$, which allows
for the possibility of bulk motion (e.g., due to \agn\ outflow or stellar 
wind).
Equation~\ref{eq:master} includes all radiation-driven and 
collision-driven processes, i.e., all
plausibly significant interactions with photons, electrons, ions, molecules, 
and dust grains and all possible excited-state decays.  It therefore provides 
a good starting point for examination of most astrophysical plasmas, excluding 
only plasmas where coherent quantum effects are important, such as 
extremely-high-density plasmas (e.g., neutron star atmospheres) or coherent 
emission from masers or lasers (discussed further in the next paragraph).
The notation is as follows: $A$ (radiative decay rate), $C$
(``collision-driven'' coefficients), $R$ (``radiation-driven'' 
coefficients), $D$ (autoionization rate), $W$ 
(grain/molecule spontaneous disintegration rate), and parentheses 
(implying notation refers to all $A$, $C$, $R$, $D$, and $W$ terms contained 
within), with the initial level as subscript and 
the final level as superscript for a particular transition between members of 
the ionic series ({\it a la} Einstein 1917).  For example, $C_{z,i}^{z',j}$ 
implies a transition from a $z$-like ion (e.g., $z=2$ for a He-like ion) in 
level $i$ to a $z'$-like ion in level $j$.  $\{nC\}_Q$ refers to the set of 
products of all collision-driven rate coefficients times their corresponding 
population densities $n$ for interactions with ions, molecules, and dust 
grains ({\em not} including electrons).  The right hand side, term-by-term,
corresponds to $(k+1)$-body recombination $(z-k,j)\rightarrow(z,i)$ (usually 
only two-body recombination is important, though three-body recombination -- 
i.e., two electrons and one ion -- is significant at high 
densities), charge 
transfer $(z-k,j)\rightarrow(z,i)$, ionization $(z,i)\rightarrow(z-k,j)$, 
excitation/deexcitation/spontaneous decay $(z,i)\rightarrow(z,j)$, 
excitation/deexcitation/spontaneous decay $(z,j)\rightarrow(z,i)$, 
$(k+1)$-body recombination $(z,i)\rightarrow(z+k,j)$, charge transfer 
$(z,i)\rightarrow(z+k,j)$, 
ionization $(z+k,j)\rightarrow(z,i)$, molecule/grain destruction $q\rightarrow q'+(z,i)$, molecule 
creation/grain adsorption $q+(z,i)\rightarrow q'$, and molecule/grain 
spontaneous disintegration $q\rightarrow q'+(z,i)$.  A similar rate equation 
to Eq.~\ref{eq:master} but for molecules or grains is straightforward to
write down as well.  In light of the simplistic approach and corresponding
assumptions adopted in
this paper, we do not investigate the coupling of Eq.~\ref{eq:master} 
with any force laws, electrodynamic equations, or thermodynamic relations.

The particular process of photoexcitation requires further elaboration.  For 
photon-atom interactions near an atomic resonance (bound-bound), there are 
two possible regimes.  One is the ``classical'' damped-oscillator regime 
(Wigner \& Weisskopf 1930; Weisskopf 1931), in which it is possible to 
separately consider the 
quantum processes of absorption (photoexcitation) and reemission driven by 
spontaneous decay.  This entire process can be thought of as 
``inelastic scattering.''  We have implicitly assumed this regime in 
our presentation of the ionic rate equations.  The alternate regime is 
resonance scattering, in which these
two processes cannot be separated out, and which always produces an outgoing 
photon with energy equal to the incident photon in the center-of-momentum frame
(``elastic scattering'').
To determine which is the valid regime requires consideration of the 
interaction timescale with the incident radiation field and the relative 
strength of the transition.  Incident monochromatic light (formally infinite 
interaction time) yields energy-preserving resonance scattering for any 
transition strength.  However, the continuous spectrum (short interaction 
time) or even absorbed power-law spectrum typical of photoionizing radiation 
and the predominance of absorption by relatively strong transitions appears 
to validate the ``classical'' regime (e.g., Sakurai 1967).  
Observationally, the difference between the 
``classical'' and resonance-scattering regimes is significant only in the 
details of the emission line profiles and in the expected degradation of 
higher-transition-energy photons to lower-transition-energy photons in the 
``classical'' regime.

\subsection{Collision-driven Processes}

The $C$ coefficients refer to all collision-driven processes, i.e., all processes 
driven by collisions of the particle of interest (usually 
an ion) with a secondary particle (usually a free electron).
Collisional excitation/ionization/deexcitation and the processes of 
recombination, charge 
transfer, and  molecule/grain creation/destruction can be compactly expressed 
as:
\begin{equation}
C_{A}^{B}=\int \big[\sigma^C\big]_{A}^{B}(E) v(E) f(E) dE,
\label{eq:collcoeff}
\end{equation}
where $A$ and $B$, for example, could be $(z,i)$ and $(z',j)$.  The integral
is over the center-of-momentum energy, $E$, of the two particles.  In the 
integrand, $\big[\sigma^C\big]_A^B(E)$ is the collisional cross 
section for the transition  $A\rightarrow B$ of the particle under 
consideration (e.g., an ion) with the interacting particle (e.g., an 
electron), $v(E)$ is the relative 
velocity of the two interacting particles, and, 
finally, $f(E)$ is the normalized energy distribution.
For isothermal electrons interacting with heavier, slower ions at the same 
temperature, $E$ is the electron kinetic energy, 
$v(E)\simeq v_{\mathrm{e}}(E)=\sqrt{2E/m_{\mathrm{e}}}$ is the 
relative velocity, and $f(E)$ is the 
electron energy distribution.  Taking the Maxwell-Boltzmann 
distribution $f(E;T)$ specified by temperature $T$ for the electron 
distribution makes the final coefficient, $C_{A}^{B}(T)$, temperature 
dependent.

\subsection{Radiation-driven Processes}

The $R$ coefficients refer to all radiation-driven processes, i.e.,
all processes driven by the local radiation field.  These coefficients 
depend solely
on the spectrum of the local radiation field and the given interaction cross
section.  Photoexcitation,
photoionization, and stimulated emission and
molecule/grain photodestruction can be expressed as:
\begin{equation}
R_{A}^{B}(F(E,\vec{r}))=\int \big[\sigma^R\big]_{A}^{B}(E,\vec{r})F(E,\vec{r}) dE,
\label{eq:radcoeff}
\end{equation}
where $A$ and $B$ again could simply be $(z,i)$ and $(z',j)$. 
The photon flux spectrum at position $\vec{r}$ is $F(E,\vec{r})$
(e.g., with in units of photons cm$^{-2}$ s$^{-1}$ keV$^{-1}$; throughout this 
paper all luminosities and fluxes will be expressed in photon units, 
{\em not} the usual energy units).  The photon flux spectrum 
may include radiation from a source external to the emission region as well as
any other sources of radiation produced in the photoionized plasma itself.  
The total radiative cross section for transition
$A\rightarrow B$ is given by $\big[\sigma^R\big]_A^B(E,\vec{r})$, which is already 
convolved with the local velocity distribution function at $\vec{r}$ of the 
particle population under consideration.  Most of the calculational
complexity of generic plasmas is reflected in $F(E,\vec{r})$, which, through
radiative transfer, is coupled to the density and velocity distributions 
(thermal or non-thermal) of all interacting particles.  However, for the 
special case of optically thin plasmas, radiative transfer is negligible,
simplifying calculations enormously (as is demonstrated below).

\section{Further Simplifications for Typical Astrophysical Plasmas}\label{sec:simp}

We now make two assumptions valid for many typical astrophysical plasmas:
(1) The only important collisional interactions involve an ion with a single
free electron, and (2) Radiation-driven and collision-driven ionizations 
remove only one electron (this does {\em not} include subsequent electron 
ejection through autoionization, which is retained in the $D$ 
coefficients).  This gives for the ground-state equation:
\begin{eqnarray}
\frac{dn_{z,g}}{dt} = \sum_j n_{z-1,j} n_{\mathrm{e}} C_{z-1,j}^{z,g}-\sum_{j} n_{z,g}
\big[R+n_{\mathrm{e}} C\big]_{z,g}^{z-1,j}-n_{z,g} \sum_j \big[R+n_{\mathrm{e}} C\big]_{z,g}^{z,j}
+\sum_{j>g}n_{z,j}A_{z,j}^{z,g}\nonumber\\
+ \sum_j n_{z+1,j} \big[R+n_{\mathrm{e}}C\big]_{z+1,j}^{z,g}+\sum_{k\geq1}\sum_{j} n_{z+k,j}D_{z+k,j}^{z,g}-n_{z,g}\sum_j n_{\mathrm{e}}C_{z,g}^{z+1,j}
\label{eq:geng}
\end{eqnarray}
and for the excited levels ($i>g$):
\begin{eqnarray}
\frac{dn_{z,i}}{dt} = \sum_j n_{z-1,j}n_{\mathrm{e}} C_{z-1,j}^{z,i}-\sum_{1\leq k< z}\sum_{j} n_{z,i} D_{z,i}^{z-k,j}-n_{z,i} \sum_{j<i} A_{z,i}^{z,j} +\sum_{j>i} n_{z,j} A_{z,j}^{z,i} + \sum_j n_{z,j}\big[R+n_{\mathrm{e}} C\big]_{z,j}^{z,i}\nonumber\\
+ \sum_j n_{z,j}\big[R+n_{\mathrm{e}} C\big]_{z,j}^{z,i}-n_{z,i} \sum_j \big[R+n_{\mathrm{e}} C\big]_{z,i}^{z,j}  +\sum_{j} n_{z+1,j}\big[R+n_{\mathrm{e}} C\big]_{z+1,j}^{z,i}+\sum_{k\geq1}\sum_{j} n_{z+k,j}D_{z+k,j}^{z,i}.
\label{eq:geni}
\end{eqnarray}

To predict the emergent spectrum, we must integrate Eqs.~\ref{eq:geng} and 
\ref{eq:geni} over the entire emission region volume.  For this purpose, we 
assume a global steady state, implying that all volume integrals 
$\int \frac{dn}{dt}\ dV$ are identically zero (i.e., we balance ionic 
population/depletion and ionic inflow/outflow for the entire volume under 
consideration).   A global steady state is assumed rather than a local steady
state, because, for example, even in a steady outflow, the local rates of 
photoionization and recombination may not necessarily balance, with 
photoionizations 
occuring nearer to the source and recombinations occuring at larger 
distances.  Even the assumption of a global steady state breaks down, of 
course, for highly-variable sources.  
However, under the global-steady-state assumption, integration of 
Eq.~\ref{eq:geng} over volume yields (with $N\equiv\int n dV$):
\begin{eqnarray}
\sum_{j>g}N_{z,j}A_{z,j}^{z,g} + \sum_{k\geq1}\sum_{j}N_{z+k,j}D_{z+k,j}^{z,g}= \int \bigg(-n_{z-1,g} n_{\mathrm{e}}
C_{z-1,g}^{z,g}+ n_{z,g} \sum_{j}\big[R+n_{\mathrm{e}} C\big]_{z,g}^{z-1,j}+n_{z,g}
\sum_j \big[R+n_{\mathrm{e}} C\big]_{z,g}^{z,j}\nonumber\\
- n_{z+1,g} \big[R+n_{\mathrm{e}}C\big]_{z+1,g}^{z,g}+n_{z,g}\sum_j n_{\mathrm{e}}C_{z,g}^{z+1,j}-n_{z,g} \sum_j \big[R+n_{\mathrm{e}} C\big]_{z,g}^{z,j}\bigg)dV
\label{eq:intgeng}
\end{eqnarray}
and integration of Eq.~\ref{eq:geni} over volume yields
\begin{eqnarray}
N_{z,i} \bigg(\sum_{j<i} A_{z,i}^{z,j} +\sum_{1\leq k< z}\sum_{j} 
 D_{z,i}^{z-k,j}\bigg)-\sum_{j>i} N_{z,j} A_{z,j}^{z,i} - \sum_{k\geq1}\sum_{j} N_{z+k,j}D_{z+k,j}^{z,i}=  \int \bigg(
 \sum_j n_{z-1,j}n_{\mathrm{e}} C_{z-1,j}^{z,i}
\nonumber\\
+ \sum_j n_{z,j}\big[R+n_{\mathrm{e}} C\big]_{z,j}^{z,i}+ \sum_j n_{z,j}\big[R+n_{\mathrm{e}} C\big]_{z,j}^{z,i}+ \sum_j n_{z+1,j}\big[R+n_{\mathrm{e}} C\big]_{z+1,j}^{z,i}-n_{z,i} \sum_j \big[R+n_{\mathrm{e}} C\big]_{z,i}^{z,j}\bigg)dV.
\label{eq:intgeni}
\end{eqnarray}
These equations are perfectly general for realistic collision- 
and/or radiation-driven plasmas.  But they are still sufficiently complicated
(due to coupling of these equations to each other and to other such equations)
to bar direct integration.

\section{Radiation-Driven Emission}\label{sec:rad}

In this section we show how further assumptions can be employed to decouple
Eqs.~\ref{eq:intgeng} and \ref{eq:intgeni} from each other and from all other 
such equations, leaving a simple analytic solution for derivation of the 
spectrum.  But first we review the atomic processes relevant for 
radiation-driven plasmas.

\subsection{Relevant Atomic Processes}\label{sec:raddrivplas}

\begin{figure*}[]
  \begin{minipage}[h]{6.5in}
\vspace{-1.1in}
    \centerline{\hspace{0in}\psfig{figure=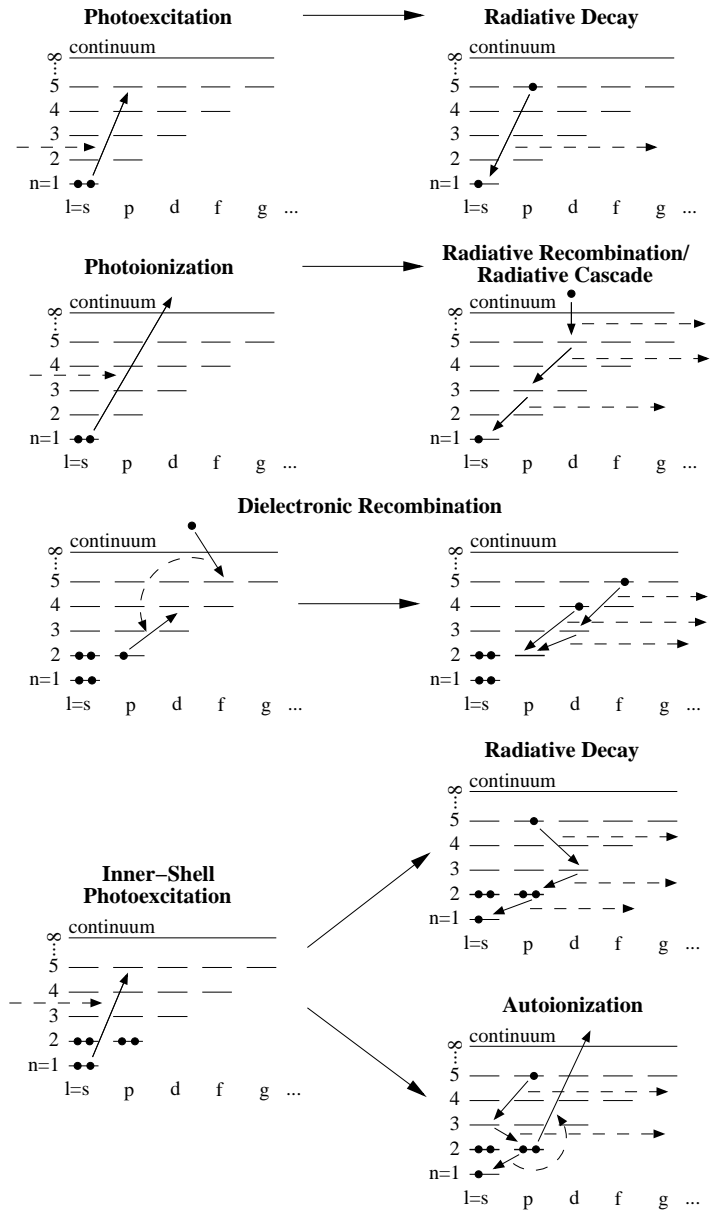,angle=-90,width=9.3in}}
\vspace{-.5in}
\caption{Highly-simplified grotrian diagrams depicting all of the relevant
atomic processes for photoionized plasmas.  The top two diagrams depict 
photoexcitation and its inverse process of radiative 
decay (only the dominant direct decay path is shown, but other decay paths 
are possible).  The next two diagrams depict photoionization and the inverse 
process of radiative recombination and radiative cascade.  
Another important inverse process 
to photoionization (especially for multi-electron ions) 
is the two-step process of dielectronic recombination, depicted
in the next two diagrams.
The last three diagrams depict photoexcitation up to a level capable of
autoionization followed by either radiative decay or autoionization; these 
diagrams can be trivially modified 
to describe photoionization to levels capable of autoionization 
as well.
\label{fig:grotrian}}
  \end{minipage}
\end{figure*}

For radiation-driven plasmas, there are only a few processes which 
contribute significantly.  These are summarized in Fig.~\ref{fig:grotrian}
and explained below.

The most important collision-driven process is recombination, including
both radiative and dielectronic recombination.
We presently discuss only radiative recombination.
Using the principle of detailed balance embodied in the Milne relation 
(Milne 1924), the radiative recombination cross section for
electrons at temperature $T$ can be related to the photoionization cross 
section:
\begin{equation}
\sigma_{z-1,g}^{z,i}(E_{\mathrm{e}})=\frac{g_{z,i}}{g_{z-1,g}}
\frac{(E_{\mathrm{th}}+E_{\mathrm{e}})^2}{m_{\mathrm{e}}c^2 E_{\mathrm{e}}}
\sigma_{z,i}^{z-1,g}(E_{\mathrm{th}}+E_{\mathrm{e}}),
\label{eq:rrsigma}
\end{equation}
where $E_{\mathrm{e}}$ is the electron kinetic energy, $E_{\mathrm{th}}$ is 
the threshold photonionizing photon energy,
$E=E_{\mathrm{th}}+E_{\mathrm{e}}$ is the photon energy, and the $g$ are the
ionic level degeneracies.  
Combining Eq.~\ref{eq:rrsigma} with Eq.~\ref{eq:collcoeff} allows for calculation
of the radiative recombination coefficients.  

The two most important radiation-driven processes
are photoexcitation and photoionization.
For the photoexcitation cross section, convolution of a line profile with 
the velocity distribution simply yields the Voigt profile (Voigt 1912):
\begin{equation}
\sigma_{z,i}^{z,j}(E)=\frac{\pi e^2}{m_{\mathrm{e}}c}f_{ji}\frac{1}{\sqrt{\pi}}\frac{1}{\Delta\nu_D}H(x,y)
\label{eq:voigt}
\end{equation}
where $\Delta\nu_D=\sqrt{2}\,\frac{\sigma_v^{\mathrm{rad}}}{c}\,\frac{E_{ji}}{h}$ and $H(x,y)$ is the Voigt function with
$x=\frac{A}{4\pi\Delta\nu_D}$ (where $A$ denotes the sum of all radiative and 
autoionizing decay rates from level $i$) and
$y=\frac{E-E_{ji}}{h\Delta\nu_D}$.  The Voigt profile is used for
both absorption and reemission line profiles (though with the possibility of
different velocity distributions due to differing bulk motion distributions).  
The photoionization cross section $\sigma_{z,i}^{z-1,g}(E)$ is 
calculated explicitly by \fac.  We convolve the photoionization cross section 
with the same $\sigma_v^{\mathrm{rad}}$ used above for the photoexcitation 
cross section.
The total cross section for photoionization and photoexcitation is taken to
be a sum of the individual cross sections.  For photoexcitation, this simply
means that we consider each transition (oscillator) decoupled from all other
transitions (oscillators): A more accurate treatment would use the full 
Kramers-Heisenberg-like cross section with appropriate radiation damping 
(e.g., Sakurai 1967), but this level of 
accuracy is rarely, if ever, needed for astrophysical sources.  The continuous 
and discrete line cross sections can therefore be taken one by one in 
Eq.~\ref{eq:radcoeff} to calculate all of the necessary
photoionization/photoexcitation coefficients.  Finally, we note that 
photoexcitation and photoionization (as well as dielectronic recombination)
can sometimes place electrons in levels capable of autoionization, as is 
illustrated in Fig.~\ref{fig:grotrian}.

\subsection{Radiation-driven Rate Equations}\label{sec:raddrivrateeq}

Returning to the ionic rate equations given in Eqs.~\ref{eq:intgeng} and 
\ref{eq:intgeni}, we now make two modifications.  First, 
introducing a powerful simplification, we assume that all ionic excited levels 
are short-lived compared to the relevant radiation-driven and collision-driven 
excitation/ionization timescales.  Second, in line with purely
radiation-driven plasmas, we assume that all collisional excitations and 
ionizations are negligible.  We therefore obtain the following 
for Eq.~\ref{eq:intgeng}:
\begin{eqnarray}
\int \bigg(n_{z,g} \sum_{j}R_{z,g}^{z-1,j}-n_{z-1,g} n_{\mathrm{e}}
C_{z-1,g}^{z,g}\bigg)dV = \sum_{k\geq1}\sum_{j}N_{z+k,j}D_{z+k,j}^{z,g} 
+ \int \bigg(n_{z+1,g} R_{z+1,g}^{z,g}-n_{z,g}\sum_j n_{\mathrm{e}}C_{z,g}^{z+1,j}\bigg)dV
\label{eq:photg}
\end{eqnarray}
and for Eq.~\ref{eq:intgeni}:
\begin{eqnarray}
N_{z,i} \bigg(\sum_{j<i} A_{z,i}^{z,j} +\sum_{1\leq k<z}\sum_{j} 
 D_{z,i}^{z-k,j}\bigg)-\sum_{j>i} N_{z,j} A_{z,j}^{z,i} - \sum_{k\geq1}\sum_{j} N_{z+k,j}D_{z+k,j}^{z,i}= \int \bigg(
 n_{z-1,g}n_{\mathrm{e}} C_{z-1,g}^{z,i} + n_{z,g}R_{z,g}^{z,i}+n_{z+1,g}R_{z+1,g}^{z,i}\bigg)dV.
\label{eq:photi}
\end{eqnarray}
These equations have a few interesting properties, which we now briefly
mention.  In
Eq.~\ref{eq:photg}, the LHS contains only those transitions which connect 
charge state $z$ to charge states with fewer 
electrons, and the RHS contains only those transitions which connect charge 
state $z$ to charge states with more electrons.  In Eq.~\ref{eq:photi}, all 
integrated expressions (involving only total ion numbers $N$) have been 
suggestively placed on the LHS with integral expressions placed on the RHS.  
The $A,R$, and $C$ coefficients in Eqs.~\ref{eq:photg} and \ref{eq:photi} 
couple the same or 
nearest-neighbor charge states.  The exception to this rule comes from the $D$
terms (representing autoionization), which, through multiple electron 
ejection, can connect more distant charge states.

Though most X-ray transitions occur very rapidly, allowing for our neglect of 
all excited-level populations, some rare long-lived levels do exist.  The most
important example is the especially long-lived $2\,^3S_1$ level in He-like 
ions (see \S\ref{sec:HHe} for more details).  The inclusion of collision- or 
radiation-driven transitions out of these long-lived levels in this
formalism requires
a much more sophisticated approach starting from Eq.~\ref{eq:intgeni}.

\subsection{Radiation-Driven Emission from H-Like and He-Like Ions}

We now apply Eqs.~\ref{eq:photg} and \ref{eq:photi} to the extremely important 
and relatively simple cases of H-like and He-like ions.  Under the assumptions 
taken so far, autoionization is unimportant {\em in}
He-like ions and {\em onto} H-like ions for the low temperatures characteristic
of photoionized plasmas (and, of course, autoionization is trivially 
non-existent {\em in} H-like ions).  We note that autoionizations, however, 
may be significant for driving ionizations {\em onto} He-like ions.  With 
these considerations in mind, we can now simplify the rate equations for all 
charge states of interest (bare, H-like, and He-like).

The absence of autoionization implies the only routes between bare, H-like, 
and He-like involve nearest neighbors (bare$\leftrightarrow$H-like and 
H-like$\leftrightarrow$He-like).  Explicitly, for 
bare$\leftrightarrow$H-like, taking $z=0$ in Eq.~\ref{eq:photg} yields:
\begin{eqnarray}
\int n_{1,g} R_{1,g}^{0}\ dV =\int n_{0}\sum_j n_{\mathrm{e}}C_{0}^{1,j}\ dV.
\label{eq:bare}
\end{eqnarray}
With a little more effort, a similar equation can also be derived for 
H-like$\leftrightarrow$He-like transitions, which together
with Eq.~\ref{eq:bare}, can both be compactly expressed as follows 
(with $z=1$ for bare$\leftrightarrow$H-like and $z=2$ for 
H-like$\leftrightarrow$He-like):
\begin{eqnarray}
\int n_{z,g} \sum_j R_{z,g}^{z-1,j}\ dV=\int n_{z-1,g}n_{\mathrm{e}} C_{\mathrm{total}}\ dV
\label{eq:photrec}
\end{eqnarray}
where $C_{\mathrm{total}}\equiv\sum_j C_{z-1,g}^{z,j}$.

Now turning to Eq.~\ref{eq:photi}, for H-like ions ($z=1$ electron), we obtain 
the following:
\begin{eqnarray}
N_{1,i} \sum_{j} A_{1,i}^{1,j}-\sum_j N_{1,j} A_{1,j}^{1,i}=\int \bigg(n_{0}
n_{\mathrm{e}} C_{0}^{1,i}+n_{1,g}R_{1,g}^{1,i}\bigg)\ dV.
\label{eq:Hi}
\end{eqnarray}
Similarly, for He-like ions ($z=2$ electrons) we obtain
\begin{eqnarray}
N_{2,i} \sum_{j<i} A_{2,i}^{2,j} -\sum_{j>i} N_{2,j} A_{2,j}^{2,i} -\sum_{k\geq1}\sum_{j} N_{2+k,j}D_{2+k,j}^{2,i}= \int \bigg(n_{1,g} n_{\mathrm{e}} C_{1,g}^{2,i} + n_{2,g}R_{2,g}^{2,i}\bigg)\ dV.
\label{eq:Hei}
\end{eqnarray}
In both cases, we omit the irrelevant final term in Eq.~\ref{eq:photi}.
The last term on the right-hand side of Eq.~\ref{eq:Hei} gives the 
contribution to the particular He-like excited level $i$ from autoionization 
of excited lower-ionization-state ions.  Autoionizations ending on He-like 
excited levels are rare compared with
those ending on the ground state (due to the significantly higher
energy change for  $nl\rightarrow1s$ 
compared to $nl\rightarrow\mathrm{n}'\mathrm{l}'$ with $\mathrm{n}'>1$).  For 
this reason, we will henceforth neglect this contribution.
Due to the symmetry of the H-like and He-like rate equations, we can 
now express Eq.~\ref{eq:photi} as:
\begin{eqnarray}
N_{z,i} \sum_{j<i} A_{z,i}^{z,j} -\sum_{j>i} N_{z,j} A_{z,j}^{z,i} = \int\bigg(
n_{z-1,g}n_{\mathrm{e}} C_{z-1,g}^{z,i}+n_{z,g} R_{z,g}^{z,i}\bigg)dV,
\label{eq:matrixtemp}
\end{eqnarray}
with (with $z=1$ for H-like and $z=2$ for He-like).  Assuming that
the temperature range is small enough that the branching ratios for 
recombination onto level $i$, $C_{z-1,g}^{z,i}/C_{\mathrm{total}}$, are 
roughly temperature-independent, we can use Eq.~\ref{eq:photrec} to reexpress 
Eq.~ \ref{eq:matrixtemp} as:
\begin{eqnarray}
N_{z,i} \sum_{j<i} A_{z,i}^{z,j} -\sum_{j>i} N_{z,j} A_{z,j}^{z,i} = 
\frac{C_{z-1,g}^{z,i}}{C_{\mathrm{total}}} \int n_{z,g}R_{z,g}^{z-1,g}\ dV+ \int n_{z,g}R_{z,g}^{z,i}\ dV.
\label{eq:HHematrix}
\end{eqnarray}
Note that the RHS is purely dependent on the total photoexcitation/photoionization 
rates out of the ground state.  For a given model of $n_{z,g}$ and all $R$, the RHS
of Eq.~\ref{eq:HHematrix} predicts the individual $N_i$ (which are direct
observables in the optically-thin limit).  Also, 
since the source terms appear linearly on the right-hand side for 
photoionization (first term) with corresponding inverse process of radiative 
recombination/radiative cascade (``REC''), and photoexcitation (second term) 
with corresponding inverse process of radiative decay (``DEC'') in 
Eq.~\ref{eq:HHematrix}, we can solve separately for each, obtaining 
$N_i=N^{\mathrm{REC}}_i+N^{\mathrm{DEC}}_i$.

\subsection{Radiative Recombination in H-like and He-like Ions}\label{sec:HHe}

H-like and He-like ions dominate the reemission spectra of photoionized 
plasmas.  Line emission in these species is also particularly simple.  
We therefore now discuss their important emission spectra in detail.

The recombination spectrum expected for a total ionic recombination rate
onto an ion with $z-1$ electrons (expressed as  $\Delta_{z-1}^{z}$), is 
determined by the following equations:
\begin{equation}
N_{z,i} \sum_{j} A_{z,i}^{z,j} -\sum_j N_{z,j} A_{z,j}^{z,i} = \Delta_{z-1}^{z}\frac{C_{z-1,g}^{z,i}}{C_{\mathrm{total}}}
\label{eq:recmatrix}
\end{equation}
with
\begin{equation}
\Delta_{z-1}^{z}=\int n_{z-1,g}n_{\mathrm{e}}\, C_{\mathrm{total}}\ dV \simeq C_{\mathrm{total}}\ \mathrm{\emm}_{z-1},
\label{eq:alpha}
\end{equation}
where $N_{z,j}$ denotes the number of ions $z$ in level $j$,
$A_{z,j}^{z,g}$ is the radiative decay rate for ion $z$ 
and transition $j\rightarrow g$ ($g$ denotes the ground state),
$C_{z-1,g}^{z,i}$ is the recombination coefficient describing recombination 
of a free electron onto ground-state ion $z-1$ creating ion $z$
with electron configuration $i$, and
$C_{\mathrm{total}}\equiv\sum_{i}C_{z-1,g}^{z,i}$.  
For simplicity, we assume that recombinations occur at 
a single representative temperature (hence the $\simeq$ in Eq.~\ref{eq:alpha}).
The emission measure EM$_{z-1}$ in Eq.~\ref{eq:alpha} is simply defined as 
$\mathrm{\emm}_{z-1}\equiv\int n_{z-1,g}n_{\mathrm{e}}\ dV$.
These equations are valid for both H-like and He-like ions (see Appendix).  
The set of equations represented by Eq.~\ref{eq:recmatrix} provide a 
complete, soluble system of equations for all levels $i>g$.  We note that the 
total population of ground-state ions is unobservable
from the plasma line emission spectrum alone; the best that can be 
done is to determine the emission measure $\mathrm{\emm}_{z-1}$.

Since we are interested in predicting line fluxes and ratios, we introduce
the line luminosity (photons~s$^{-1}$):
\begin{eqnarray}
L_{j\rightarrow i}=N_j A_{z,j}^{z,i}=\Delta_{z-1}^{z} l_{j\rightarrow i},
\end{eqnarray}
where $l_{j\rightarrow i}=N_j A_{z,j}^{z,i}/\Delta_{z-1}^{z}$ is the dimensionless line 
luminosity coefficient.  Similarly, the recombination luminosity onto level 
$i$ is
\begin{eqnarray}
L_{f\rightarrow i}=\Delta_{z-1}^{z} \frac{C_{z-1,g}^{z,i}}{C_{\mathrm{total}}}=\Delta_{z-1}^{z} l_{f\rightarrow i},
\end{eqnarray}
where $l_{f\rightarrow i}= C_{z-1,g}^{z,i}/C_{\mathrm{total}}$ is the 
recombination luminosity coefficient (The subscript $f$ stands for the 
initial state with a ``free'' electron).

\begin{figure*}[]
   \begin{minipage}[h]{6.5in}
     \centerline{\hspace{0.27in}\psfig{figure=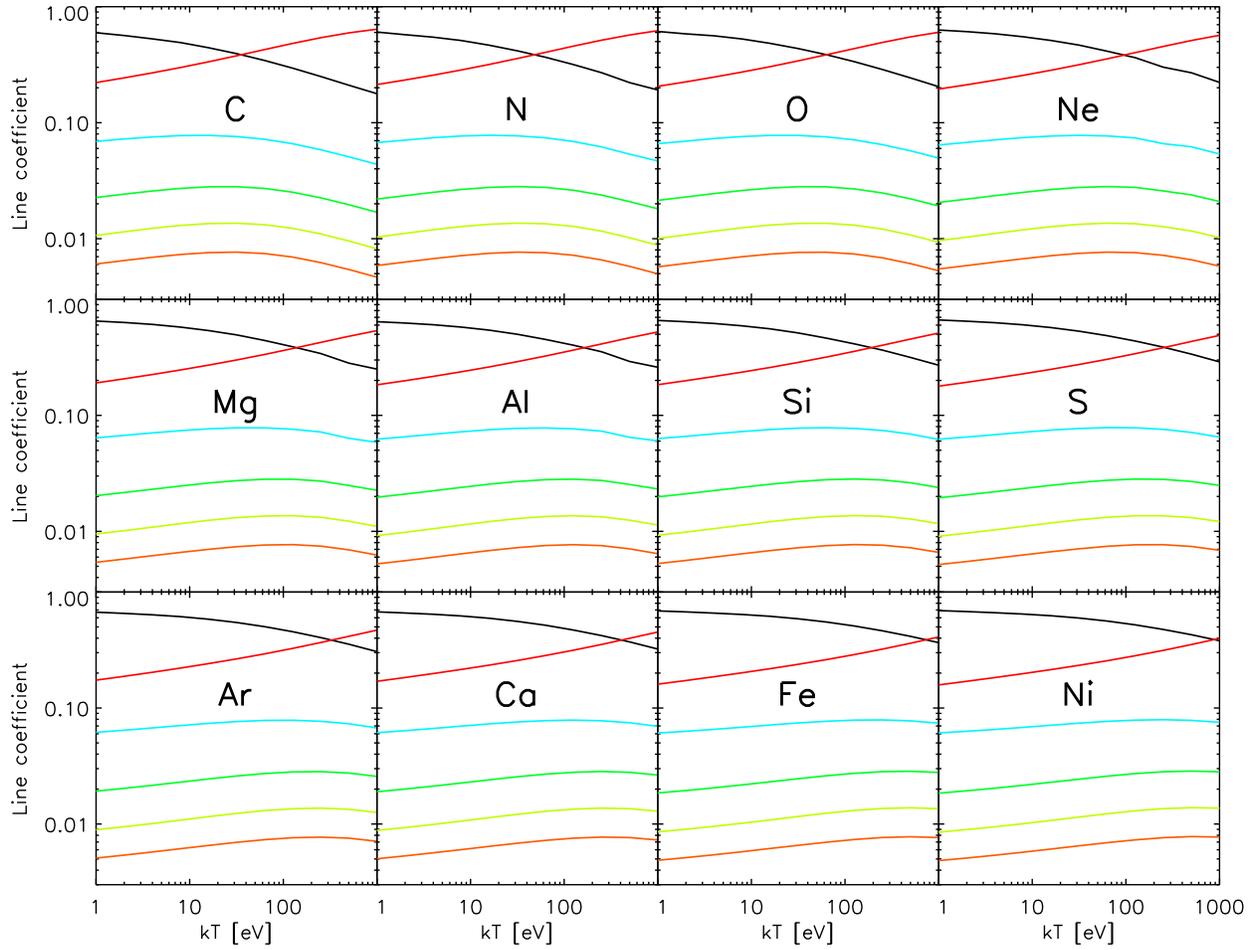,angle=90,width=7in}}
\vspace{-.1in}
\caption{Line coefficients for the brightest features resulting
from radiative recombination forming H-like ions.  Ly~$\alpha$ (black), 
Ly~$\beta$ (cyan), Ly~$\gamma$ (green), Ly~$\delta$ (olive), and 
Ly~$\epsilon$ (orange) are shown, along with the ground-state \rrc\ (red).  
Division of the observed line luminosity by the line 
coefficient gives the radiative recombination rate $\Delta_{z-1}^{z}$ (at an assumed 
temperature).  Division of $\Delta_{z-1}^{z}$ by $C_{\mathrm{total}}$ in 
Fig.~\ref{fig:rec_h_rate} gives the ionic emission measure 
\emm$_{z-1}=\int n_{\mathrm{z-1,g}} n_{\mathrm{e}} dV$ 
(see Eq.~\ref{eq:alpha}).
\label{fig:rec_h}}
   \end{minipage}
\end{figure*}

\begin{figure*}[]
   \begin{minipage}[h]{6.5in}
     \centerline{\hspace{0.37in}\psfig{figure=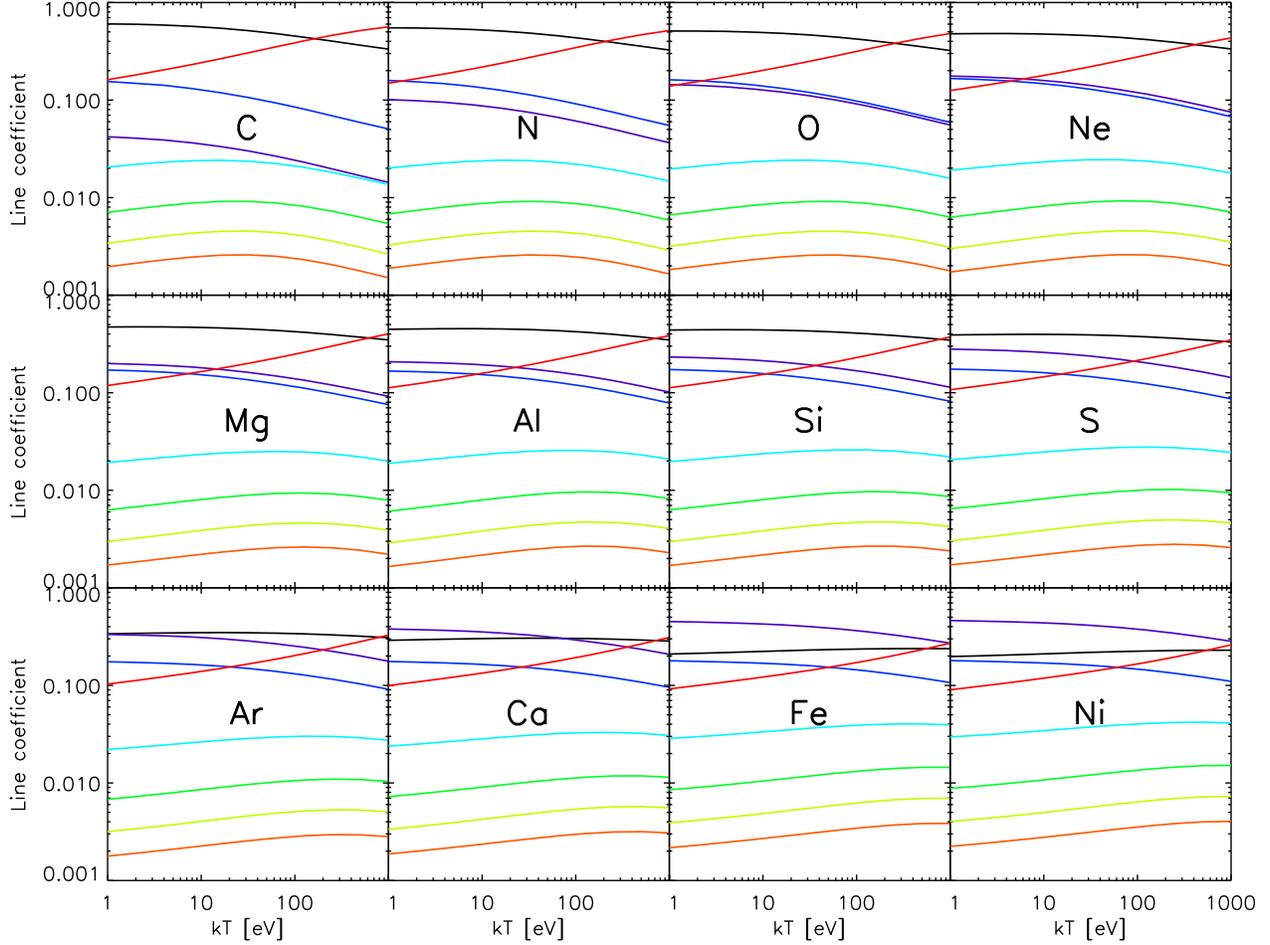,angle=90,width=7in}}\vspace{-0.1in}
\caption{Line coefficients for the brightest features resulting
from pure radiative recombination forming He-like ions in a low density plasma.
Forbidden (black), intercombination (purple), resonance (blue), 
He$\beta$ (cyan), He$\gamma$ (green), He$\delta$ (olive), and He$\epsilon$ 
(orange) are shown as well as the ground-state \rrc\ (red).  
Division of the observed line luminosity by the line 
coefficient gives the radiative recombination rate $\Delta_{z-1}^{z}$.  Division of 
$\Delta_{z-1}^{z}$ by $C_{\mathrm{total}}$ in Fig.~\ref{fig:rec_he_rate} gives the 
ionic emission measure \emm$_{z-1}=\int n_{\mathrm{z-1,g}} n_{\mathrm{e}} dV$ 
(see Eq.~\ref{eq:alpha}).  Caveat: At the low temperature end, only 
radiative recombination is important, but at the high temperature end 
($kT\sim E_0$ where $E_0$ is the resonance line energy), additional 
contributions to the line emission (not included) from dielectronic 
recombination and collisional excitation become important.  Also, high
densities or strong ambient UV fields can convert $f\rightarrow i$,
as explained in \S\ref{sec:HHe}.
\label{fig:rec_he}}
   \end{minipage}
\end{figure*}

\begin{figure*}[]
   \begin{minipage}[h]{6.5in}
     \centerline{\hspace{0in}\psfig{figure=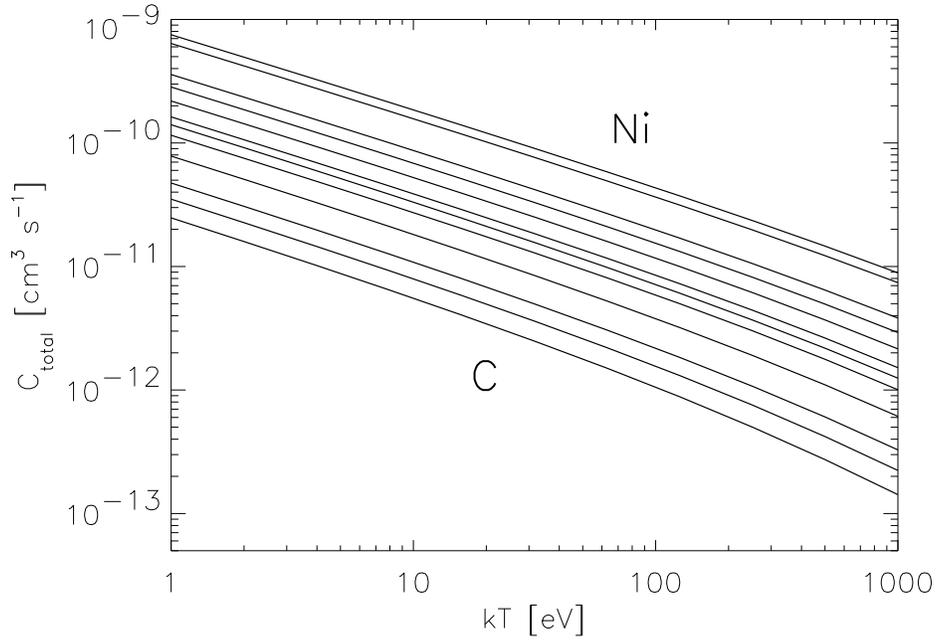,angle=90,width=5in}}
\vspace{-.2in}
\caption{The temperature dependence of $C_{\mathrm{total}}$ for radiative
recombination forming H-like ions (see also Fig.~\ref{fig:rec_h}).
\label{fig:rec_h_rate}}
   \end{minipage}
\end{figure*}

\begin{figure*}[]
   \begin{minipage}[h]{6.5in}
     \centerline{\hspace{0in}\psfig{figure=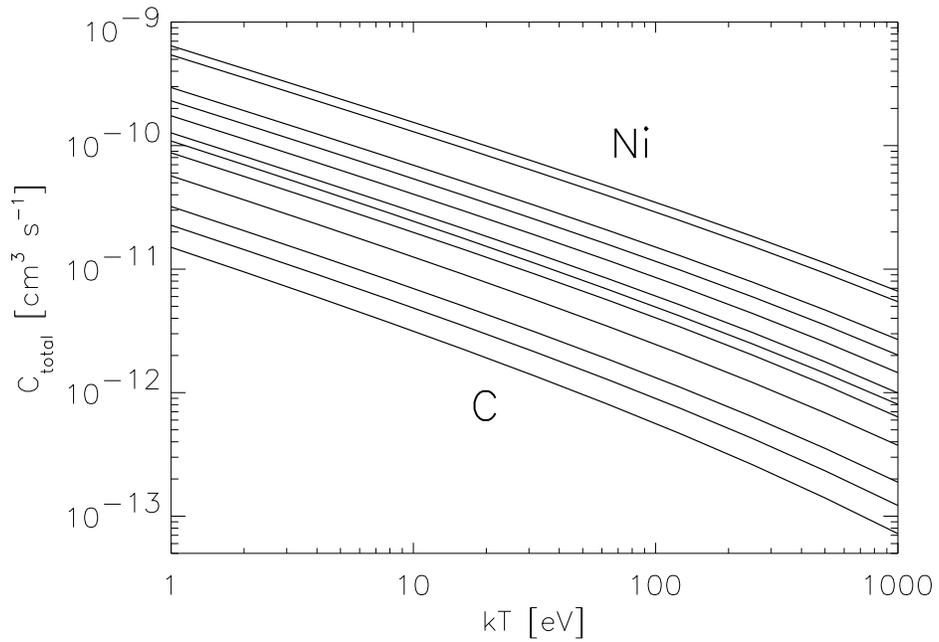,angle=90,width=5in}}
\vspace{-0.2in}
\caption{The temperature dependence of $C_{\mathrm{total}}$ for radiative
recombination forming He-like ions (see also Fig.~\ref{fig:rec_he}).  
Dielectronic recombination, which
becomes important at the high temperature end ($kT\sim E_0$ where 
$E_0$ is the resonance line energy), has {\em not} been included.
\label{fig:rec_he_rate}}
   \end{minipage}
\end{figure*}

\begin{figure*}[]
   \begin{minipage}[h]{6.5in}
     \centerline{\hspace{0.25in}\psfig{figure=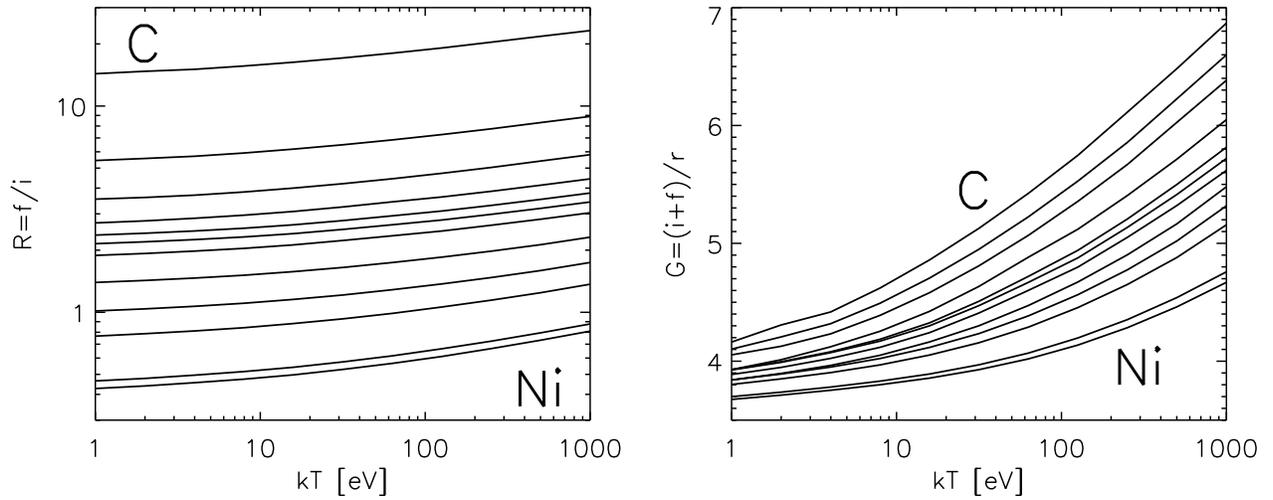,angle=90,width=7.4in}}
\vspace{-1.3in}
\caption{He-like triplet ratios $R=f/i$ and $G=(i+f)/r$ for 
pure radiative recombination for elements C, N, O, Ne, Mg, Al, Si, S, Ar, Ca, 
Fe, and Ni (from top to bottom).  At the low temperature end, only radiative
recombination is important, but at higher temperatures ($kT\sim E_0$ where 
$E_0$ is the resonance line energy), additional contributions (not included) 
from dielectronic recombination and collisional excitation become important.
\label{fig:RG}}
   \end{minipage}
\end{figure*}

In Figs.~\ref{fig:rec_h} and 
\ref{fig:rec_he}, we plot the line coefficients for the principal lines
of H-like and He-like ions, respectively.
Also shown in Figs.~\ref{fig:rec_h_rate} and \ref{fig:rec_he_rate}
are the corresponding $C_{\mathrm{total}}$ as
a function of temperature.  Together, these plots allow for the determination of the
total recombination rates as well as ionic emission measures for a given photoionized
plasma spectrum. 
In addition, we provide the standard He-like triplet
ratios $R=f/i$ and $G=(i+f)/r$ (Gabriel \& Jordan 1969) for pure radiative 
recombination as a function of temperature in Fig.~\ref{fig:RG}.  We stress
that since the He-like curves do not include the 
additional contributions from dielectronic recombination and collisional 
excitation present at the high temperature end, direct comparison with data
for these high temperatures should be avoided.  Explanation of exactly how these 
calculations were carried out is provided below in \S\ref{sec:atom}.

We again stress the neglect of the well-observed process of 
forbidden-to-intercombination line conversion in the He-like triplets through 
excitation of the long-lived 2 $^3$S$_1$ level up to the 2 $^3$P multiplet
(Gabriel \& Jordan 1969).  Though we do not take this into account here 
(and it is {\em not} included in \photoion), it is simple enough to 
determine the effects on the He-like triplet ratio $R=f/i$ of electron collisional 
excitation which is dependent on $n_{\mathrm{e}}$ and $T_{\mathrm{e}}$ (e.g., 
Porquet \& Dubau 2000) and/or of UV photoexcitation which is dependent only on 
$F(E_0)$ (e.g., Kahn \etal\ 2001).

\subsection{Radiation-Driven Emission from Other Ions}

Li-like and lower-ionization-state ions are in general more complicated
for a number of reasons.  Excitations and ionizations can now
lead to excited levels capable of autoionization of
single and, importantly, multiple electrons.  This implies that the strict 
equivalence between the single-electron removal rate (either through 
photoionization or single-electron autoioinization following 
photoexcitation) and the recombination rate between neighboring charge 
states are no longer necessarily equal.  However, autoionizations removing
more than one electron are typically negligible compared to 
photoionizations and single-electron autoionization (though this statement
deserves further investigation).  Therefore, this
equivalence is likely still a very good approximation, and can be expressed
as follows:
\begin{eqnarray}
\Delta_{z-1}^{z}= \int n_{z-1,g}n_{\mathrm{e}}C_{\mathrm{total}}\ dV=\int n_{z,g} \bigg(\sum_j R_{z,g}^{z-1,j}+ \sum_j R_{z,g}^{z,j}f^D_{z,j}\bigg)dV,
\label{eq:otherrec}
\end{eqnarray}
where $C_{\mathrm{total}}$ is the total recombination rate coefficient (now
for both radiative and dielectronic recombination) and
$f^D_{z,j}$ denotes the fraction of decays from $z,j$
that lead to electron removal via autoionization (hence the $D$ superscript).

We now focus on Eq.~\ref{eq:photi}, which describes the line emission. 
As a further approximation, we drop the 
last term on the RHS of
Eq.~\ref{eq:photi}, which gives all autoionizations from lower charge states 
landing on the excited level $z,i$.  This allows us to consider transitions 
only within the same 
ion.  As for Eq.~\ref{eq:HHematrix}, we assume that the temperature range is 
small enough so that $C_{z-1,g}^{z,i}/C_{\mathrm{total}}$ is independent of 
temperature.  Then, from Eqs.~\ref{eq:photi} and \ref{eq:otherrec}, we obtain:
\begin{eqnarray}
N_{z,i} \bigg(\sum_{j<i} A_{z,i}^{z,j}+\sum_{1\leq k< z}\sum_{j}
 D_{z,i}^{z-k,j}\bigg) -\sum_{j>i} N_{z,j} A_{z,j}^{z,i} \approx\frac{C_{z-1,g}^{z,i}}{C_{\mathrm{total}}}\int
 n_{z,g}\bigg(\sum_j R_{z,g}^{z-1,j}+ \sum_j R_{z,g}^{z,j}f^D_{z,j}\bigg)dV \nonumber\\
+\int n_{z,g}R_{z,g}^{z,i}dV+\int n_{z+1,g}R_{z+1,g}^{z,i}dV.
\label{eq:otherfinali}
\end{eqnarray}
The linearity of Eq.~\ref{eq:otherfinali} allows us to separate out the 
respective contributions to all $N_{z,i}$ from photoionization/autoionization
(eventually producing recombinations, both radiative and dielectronic) and 
the fraction of photoexcitations that produce radiative decays:
$N_{z,i}=N^{\mathrm{REC}}_{z,i}+N^{\mathrm{DEC}}_{z,i}$.  
For photoionization/autoionization which produce eventual recombinations, we 
have:
\begin{eqnarray}
N^{\mathrm{REC}}_{z,i} \sum_{j<i} A_{z,i}^{z,j} -\sum_{j>i} N^{\mathrm{REC}}_{z,j} A_{z,j}^{z,i} \approx \frac{C_{z-1,g}^{z,i}}{C_{\mathrm{total}}}\int n_{z,g}\bigg(\sum_j R_{z,g}^{z-1,j}+ \sum_j R_{z,g}^{z,j}f^D_{z,j}\bigg)dV.
\label{eq:fe_rec}
\end{eqnarray}
We have neglected the $\sum_{1\leq k< z}\sum_{j} D_{z,i}^{z-k,j}$ terms,
which for the low temperatures typical of photoionized plasmas do not usually 
contribute significantly to the recombination spectra.  These terms are 
retained, however, in Eq.~\ref{eq:fe_dec}, where they are essential for 
excitations and ionizations resulting in levels capable of autoionization.
The system of equations represented by Eq.~\ref{eq:fe_rec} can be thought of 
as the exact solution of the line spectrum for a given 
recombination rate onto each level $i$ of 
$\Delta_{z-1}^{z} C_{z-1,g}^{z,i}/C_{\mathrm{total}}$, where $\Delta_{z-1}^{z}$ is the total 
recombination rate.
Similarly, for photoexcitation, we obtain
\begin{eqnarray}
N^{\mathrm{DEC}}_{z,i} \bigg(\sum_{j<i} A_{z,i}^{z,j}+\sum_{1\leq k< z}\sum_{j}
 D_{z,i}^{z-k,j}\bigg) -\sum_{j>i} N^{\mathrm{DEC}}_{z,j} A_{z,j}^{z,i} \approx \int n_{z,g}R_{z,g}^{z,i}dV+\int n_{z+1,g}R_{z+1,g}^{z,i}dV.
\label{eq:fe_dec}
\end{eqnarray}
Here, for the first time on the LHS, we must integrate over both $n_{z,g}$ and 
$n_{z+1,g}$, the latter represent all photoionizations that 
leave the ion in an excited level (for photoionization of H- and 
He-like ions, this is always the ground state).  As a further
simplification to Eq.~\ref{eq:fe_dec}, we can discard the last term on the 
LHS, which describes feeding of $z,i$ by all higher excitation levels:
\begin{eqnarray}
N^{\mathrm{DEC}}_{z,i} \bigg(\sum_{j<i} A_{z,i}^{z,j}+\sum_{1\leq k< z}\sum_{j}
 D_{z,i}^{z-k,j}\bigg) \approx \int n_{z,g}R_{z,g}^{z,i}dV+\int n_{z+1,g}R_{z+1,g}^{z,i}dV.
\label{eq:fe_dec_approx}
\end{eqnarray}
This directly gives the fluorescence yield for each excited level without
having to perform a matrix inversion.

\subsection{Atomic Data Calculation and Implementation}\label{sec:atom}

In the foregoing, we have delineated how a series of approximations can
reduce solution of the spectrum of each ion to a seemingly simple matrix 
inversion.
We now discuss how these matrix inversions are currently implemented in
\photoion.

Throughout, we use atomic data generated by the atomic code FAC 
(Gu 2002; Gu 2003).   For recombination calculations onto 
all relevant ions (for which recombination produces X-ray features), we 
explicitly calculate all levels up to principal 
quantum number $n\leq 25$ and all E1, E2, M1, and M2 transitions 
$n'l'\rightarrow nl$ with $n,n'\leq 25$.  Explicit calculation of all levels 
for $n=45$ allows for interpolation for levels $n=26$--44.  Using the H-like 
approximation for the recombination cross-sections, we sum all recombinations 
onto levels $n>45$ and 
then spread this contribution over $n=45$.  Solution of the
relevant matrices (such as in Eqs.~\ref{eq:HHematrix} and \ref{eq:recmatrix}) 
can now be obtained using only levels with $n\leq45$.
We also include all important two-photon decays for both H-like and He-like 
(Drake 1986); such decays are not calculated by \fac.
Similar calculations were carried out to determine the contribution from 
radiative and dielectronic recombination forming L-shell ions for all
ions of interest in the X-ray regime.  Due to the large number of excited
levels for L-shell ions, inversions of the relevant matrix 
(represented by Eq.~\ref{eq:fe_rec}) required further approximations described 
in Gu (2002) and Gu (2003).

Photoexcitation to all $n\leq25$ levels ($n\leq100$ for H- and He-like ions) 
is calculated directly from the atomic code results for the oscillator 
strengths $f_{ji}$ and radiative decay rates $A_{ji}$ using the standard 
Voigt profile for line absorption/emission discussed in 
\S\ref{sec:raddrivplas}.  (Caveat: Since we do not presently calculate 
photoexcitation to higher $n$ levels --- ideally out to $n\rightarrow\infty$ 
--- there is an artificial discontinuity at the photoelectric edges.)
For radiative decay following valence-shell photoexcitation, solving the 
matrix for all excited levels is unnecessary since the dominant decay is the 
direct decay back to the ground state.  Currently, to simplify and speed up 
the code, we therefore assume that every
photoexcitation always produces a transition back to the ground 
state (even for photoexcitations to levels capable of autoionization, see
below).  This approximation is reliable for all important valence-shell
transitions at the 
few percent level for reemission in an optically thin medium.  For the same
reason, we also only use the radiative decay rate to the ground for parameter 
$x$ in the Voigt function (Eq.~\ref{eq:voigt}).  Velocity widths in many X-ray
absorbers have recently been observed to be significantly larger than the 
expected thermal width, which itself is much larger than the natural line 
width, making this a safe approximation.

To increase the accuracy for especially important transitions (currently
only in the H- and He-like ions), we verified 
and, if necessary, modified important photoelectric cross sections, oscillator 
strengths, and especially wavelengths with previously-published calculations 
and experimental results from Verner \etal\ (1996); 
Verner, Verner, \& Ferland (1996); and NIST 
\footnote{http://physics.nist.gov/cgi-bin/AtData/main\_asd}.

Also, not yet included is the 
calculation of all possible autoionization rates following 
photoexcitation or photoionization to determine total resulting line 
fluorescence.  Currently, all photoexcitations
are assumed to decay directly back to the ground state, leading to a
sometimes gross overprediction of fluorescence for some excited levels
with especially high autoionization rates.
Oppositely, fluorescence resulting from photoionization is currently absent 
in the code.  Prospects for inclusion of these rates are discussed in 
\S\ref{sec:dis}.

\section{Irradiated Cone Model}\label{sec:cone}

\begin{figure*}[]
  \begin{minipage}[h]{6.5in}	
\vspace{-0.9in}
    \centerline{\hspace{.93in}\psfig{figure=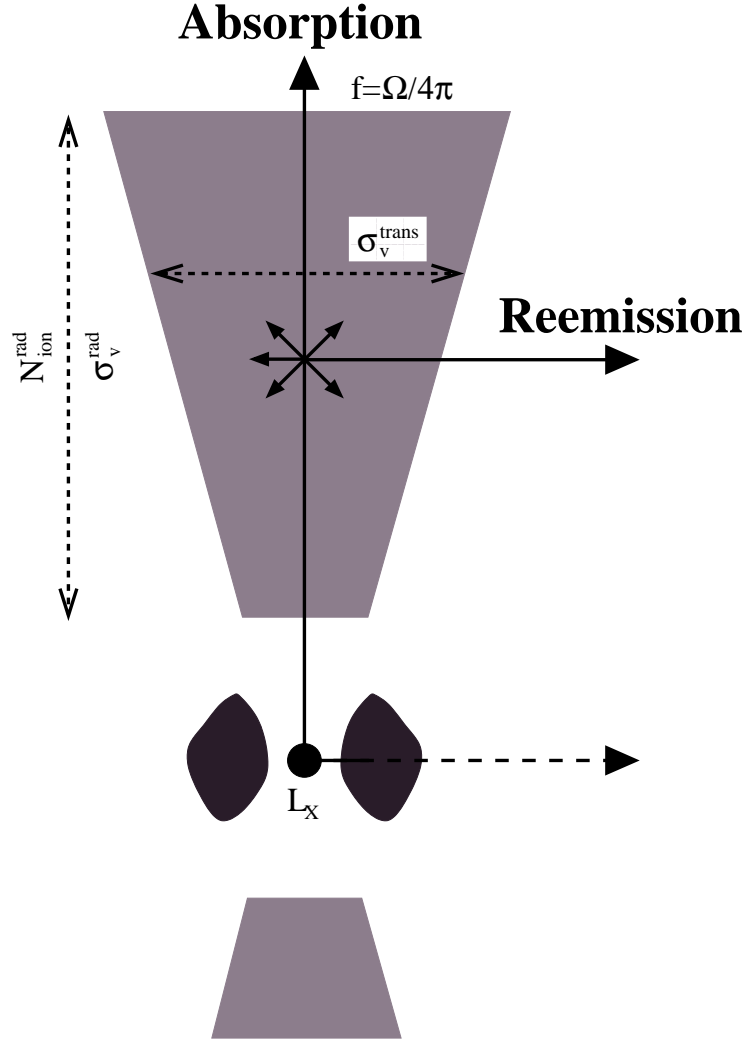,angle=0,width=6.0in}}
\vspace{-1.2in}
\caption{Simple cartoon of irradiated cone model (not to scale).  
The central nuclear component which irradiates the cone is shown as the black 
spot.  The labels ``Absorption'' or ``Reemission'' denote the two 
main orientations of interest.  
The following model parameters are indicated: $f=\Omega/4\pi$ (covering 
factor), $N^{\mathrm{rad}}_{\mathrm{ion}}$ (each individual ionic 
column density), $\sigma_v^{\mathrm{rad}}$ (radial gaussian velocity 
width), $\sigma_v^{\mathrm{trans}}$ (transverse gaussian velocity 
width), and $L_X$ (irradiating spectrum).
In the ``Absorption'' direction, in addition to features imprinted 
by line and edge absorption, reemission in the cone may also provide an 
additional contribution (see Fig.~\ref{fig:sy1}).  Also, in the ``Reemission'' 
direction, the intrinsic continuum may or may not be obscured by an 
additional medium (see Fig.~\ref{fig:sy2}).  Finally, we note 
that some sources may also exhibit a second cone (shown at the bottom of the 
figure).  
    \label{fig:cone}}
  \end{minipage}
\end{figure*}

In order to integrate the expressions on the right-hand sides of 
Eqs.~\ref{eq:HHematrix} and \ref{eq:otherfinali}, we must
first introduce our model.  The simplest version consists 
of an isotropic cone irradiated by a nuclear source with photon luminosity 
spectrum $L(E)= AE^{-\Gamma}$ (e.g., in units photons~s$^{-1}$~keV$^{-1}$).  
We characterize the cone by a solid angle $\Omega$, radial column density $G$, 
and radial gaussian velocity distribution $\sigma_v^{\mathrm{rad}}$ (which 
broadens all radiative cross-sections) for the ground state of each 
particular ion (Fig.~\ref{fig:cone}).  We first consider the more general 
non-isotropic case: $L(E,\vec{\theta})$, $G(\vec{\theta})$,  and 
$\sigma_v^{\mathrm{rad}}(\vec{\theta})$, where 
$\vec{\theta}\equiv(\theta,\phi)$.

Employing the above assumptions, we obtain for the $R$ 
coefficients pertaining to a particular ion $n_{z,i}$ (with $z'=z$ for 
photoexcitation and $z'=z-1$ for photoionization):
\begin{equation}
R_{z,i}^{z',j}(F(E,r,\vec{\theta}))=\int \big[\sigma^R\big]_{z,i}^{z',j}(E,\vec{\theta})
\bigg(\frac{L(E,\vec{\theta})}{4\pi r^2} e^{-\tau(E,r,\vec{\theta})}\bigg) dE,
\end{equation}
where $\tau(E,r,\vec{\theta})=\int_0^r \sum_{\{Z,z\}}{n_{Z,z,g}}(r',\vec{\theta})\sigma^R_{Z,z}(E,\vec{\theta})dr'$ and $\sigma^R_{Z,z}(E,\vec{\theta})$ denotes the total 
radiative cross section for all radiation-driven transitions from 
the ground state of a particular element $A$ with $z$ electrons, with all cross 
sections already implicitly convolved with a possibly angular-dependent 
velocity distribution along the cone, $\sigma^{\mathrm{rad}}_v(\vec{\theta})$.
$L(E,\vec{\theta})$ is the intrinsic nuclear luminosity (again with possible 
angular dependence).  The exponential takes into account absorption along 
the cone due to the ground-state number densities $n_{z,g}$.

We now integrate the $R$ coefficients times the appropriate
ionic density over volume to get the total 
photoexcitation (or photoionization) rate in the plasma:
\begin{equation}
\int n_{z,g}(r,\vec{\theta})R_{z,g}^{z',j}(F(E,r,\vec{\theta}))dV=\int dE\int \frac{d\Omega}{4\pi}\int dr\ n_{z,g}(r,\vec{\theta})\big[\sigma^R\big]_{z,g}^{z',j}(E,\vec{\theta})\ L(E,\vec{\theta}) e^{-\tau(E,r,\vec{\theta})}.
\label{eq:intRvolume}
\end{equation}

There are two different assumptions which can be made at this point to
simplify this integral.  The first assumption is that only the particular
ion of interest contributes to the opacity.  This is fairly robust
if column densities of all other ions are low enough that only discrete line 
absorption is present in the spectrum, since the opacity due to lines alone is 
usually negligible.  Under this assumption, the opacity $\tau$ is only 
dependent on the particular ion of 
interest; therefore,
$d\tau\equiv d\tau(E,r,\vec{\theta})= n_{z,g}(r,\vec{\theta})\sigma^R_z(E,\vec{\theta})dr$, and we can rewrite Eq.~\ref{eq:intRvolume} as:
\begin{equation}
\int n_{z,g}(r,\vec{\theta})R_{z,g}^{z',j}(F(E,r,\vec{\theta}))dV=\int dE\int \frac{d\Omega}{4\pi}\int_0^{\tau} d\tau'\ L(E,\vec{\theta}) e^{-\tau'}.
\end{equation}
Upon integration over $\tau'$, this simplifies to:
\begin{eqnarray}
\int n_{z,g} R_{z,g}^{z',j}dV=\int
\frac{d\Omega}{4\pi}G_{z}(\vec{\theta})\int dE\
\big[\sigma^R\big]_{z,g}^{z',j}(E,\vec{\theta})\ L(E,\vec{\theta})\frac{1-e^{-\tau(E,\vec{\theta})}}{\tau(E,\vec{\theta})},
\label{eq:Gthetaone}
\end{eqnarray}
where $G_{z}$ is the ground-state column density for ions (with $z$ electrons)
of the particular element under consideration.

Alternatively, a second possible assumption could be that relative ionic 
abundances are independent of radius.  Starting once
again from Eq.~\ref{eq:intRvolume}, we now simply multiply and divide by the 
sum in large parentheses in the following expression:
\begin{equation}
\int n_{z,g}(r,\vec{\theta})R_{z,g}^{z',j}(F(E,r,\vec{\theta}))dV=\int dE\int \frac{d\Omega}{4\pi}\int dr\ \frac{n_{z,g}(r,\vec{\theta})\big[\sigma^R\big]_{z,g}^{z',j}(E,\vec{\theta})}{\sum_{\{Z',z\}}{n_{Z',z,g}}(r,\vec{\theta})\sigma^R_{Z',z}(E,\vec{\theta})}\bigg(\sum_{\{Z',z\}}n_{Z',z,g}(r,\vec{\theta})\sigma^R_{Z',z}(E,\vec{\theta})\bigg)\ L(E,\vec{\theta}) e^{-\tau(E,r,\vec{\theta})}.
\label{eq:fractionalterm}
\end{equation}
Recall that $d\tau(E,r,\vec{\theta})=\sum_{\{Z,z\}}{n_{Z,z,g}}(r,\vec{\theta})\sigma^R_{Z,z}(E,\vec{\theta})dr$.  Therefore, the sum in large parentheses in Eq.~\ref{eq:fractionalterm} 
times $dr$ is just $d\tau$.  Therefore, by our central assumption that all
relative ionic abundances are independent
of $r$, the fractional term in Eq.~\ref{eq:fractionalterm} is also 
independent of $r$ and can be replaced by a fraction involving only
the total integrated ionic column densities $G_{Z,z}$:
\begin{equation}
\int n_{z,g}(r,\vec{\theta})R_{z,g}^{z',j}(F(E,r,\vec{\theta}))dV=\int\frac{d\Omega}{4\pi}\int dE  \frac{G_{z}(\vec{\theta})\big[\sigma^R\big]_{z,g}^{z',j}(E,\vec{\theta})}{\sum_{\{Z,z\}}{G_{Z,z}}(\vec{\theta})\sigma^R_{Z,z}(E,\vec{\theta})}\ L(E,\vec{\theta}) \int_0^{\tau} e^{-\tau'}d\tau'.
\end{equation}
Upon integration over $\tau$, we obtain:
\begin{eqnarray}
\int n_{z,g} R_{z,g}^{z',j}dV=\int
\frac{d\Omega}{4\pi}G_{z}(\vec{\theta})\int dE\
\big[\sigma^R\big]_{z,g}^{z',j}(E,\vec{\theta})\ L(E,\vec{\theta})\frac{1-e^{-\tau(E,\vec{\theta})}}{\tau(E,\vec{\theta})},
\label{eq:Gthetatwo}
\end{eqnarray}
which, interestingly, is identical to Eq.~\ref{eq:Gthetaone}.

For the simplest model of an isotropic cone
of material with solid angle extent $\Omega$ characterized 
by angle-independent radial column density $G$, radial gaussian velocity 
distribution $\sigma_v$, and nuclear photon luminosity spectrum $L(E)$, the 
identical expressions derived above (Eqs.~\ref{eq:Gthetaone} and 
\ref{eq:Gthetatwo}) reduce to:
\begin{equation}
\int n_{z,g} R_{z,g}^{z',j}dV=\frac{\Omega}{4\pi}G_{z}\int dE\ \big[\sigma^R\big]_{z,g}^{z',j}(E)\ L(E)\frac{1-e^{-\tau(E)}}{\tau(E)}.
\label{eq:photoion}
\end{equation}
This equation serves as the basis of \photoion.

\section{Spectra for Various Absorption and Reemission Geometries Calculated with \photoion}\label{sec:examples}

The {\small XSPEC} model \photoion\ allows the calculation of spectra arising
from different absorption and reemission geometries.  Below, we provide 
examples of some simple geometries possible with \photoion.

\subsection{Absorption}\label{sec:absorption}

\begin{figure*}[]
  \begin{minipage}[h]{6.5in}
    \vspace{-.2in}
    \centerline{\hspace{.55in}\psfig{figure=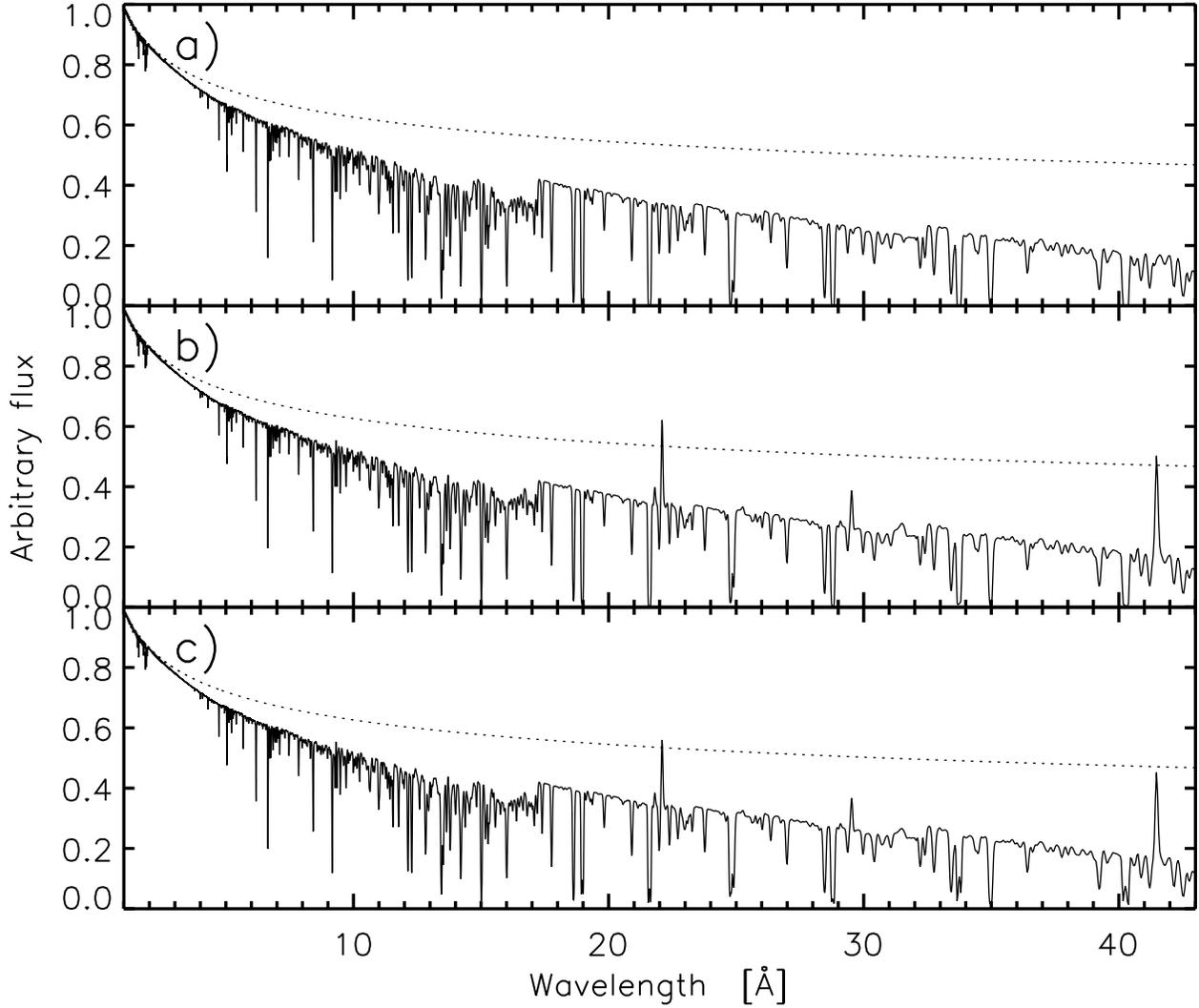,angle=90,width=8.4in}}
    \vspace{-0.45in}
\caption{Three examples of multi-phase ``Absorption'' spectra 
(see Fig.~\ref{fig:cone}) calculated by \photoion:  a) Pure absorption, 
b) Absorption plus reemission (upper limit assuming $\mathrm{Opacity}(E)=1$, 
see \S\ref{sec:absreem}), 
and c) Absorption plus reemission (lower limit assuming 
$\mathrm{Opacity}(E)=\mathrm{e}^{-\tau(E)}$, see \S\ref{sec:absreem}).  
For all spectra, we take $\sigma_v^{\mathrm{rad}}=400$~km~s$^{-1}$ 
and a power-law slope of $\Gamma=1.8$.  And, for spectra in panels b) and c), 
we further assume a covering fraction of $f=\Omega/4\pi=0.15$ 
for the reemission component.
\label{fig:sy1}}
  \end{minipage}
\end{figure*}

An example of a spectrum due to pure absorption is given in Fig.~\ref{fig:sy1}a.
The imprinting of absorption features on the intrinsic continuum is
determined by $e^{-\tau(E)}$ with
\begin{eqnarray}
\tau(E)\equiv\tau(E,\vec{\theta})=\int_0^r\sum_{\{Z,z\}}{n_{Z,z,g}}(r',\vec{\theta})\sigma^R_{Z,z}(E,\vec{\theta})dr'=\sum_{\{Z,z\}}\int_0^r {n_{Z,z,g}}(r',\vec{\theta})\sigma^R_{Z,z}(E,\vec{\theta})dr'=\sum_{\{Z,z\}}G_{Z,z}\sigma^R_{Z,z}(E,\vec{\theta}).
\label{eq:abs}
\end{eqnarray}
Here, $\sigma^R_{Z,z}(E,\vec{\theta})$
denotes the total radiative cross section (in the line-of-sight direction 
$\vec{\theta}$) for all radiation-driven transitions from 
the ground state of each ion, with all cross-sections implicitly convolved with
the velocity distribution characterized by width $\sigma_v^{\mathrm{rad}}$.  
This model can also be calculated using the related multiplicative \xspec\ model 
{\small MPABS} (multi-phase absorber), which allows for pure absorption studies through 
the independent variation of each ionic column density.  These models should be useful 
in particular for understanding the complex, multi-phase absorption observed in a 
number of Seyfert~1 galaxy spectra (e.g., Kaastra \etal\ 2000, Sako \etal\ 2001, 
Behar, Sako, \& Kahn 2001).

\subsection{Reemission}\label{sec:reemission}

\begin{figure*}[]
  \begin{minipage}[h]{6.5in}
    \vspace{-0.2in}
    \centerline{\hspace{.55in}\psfig{figure=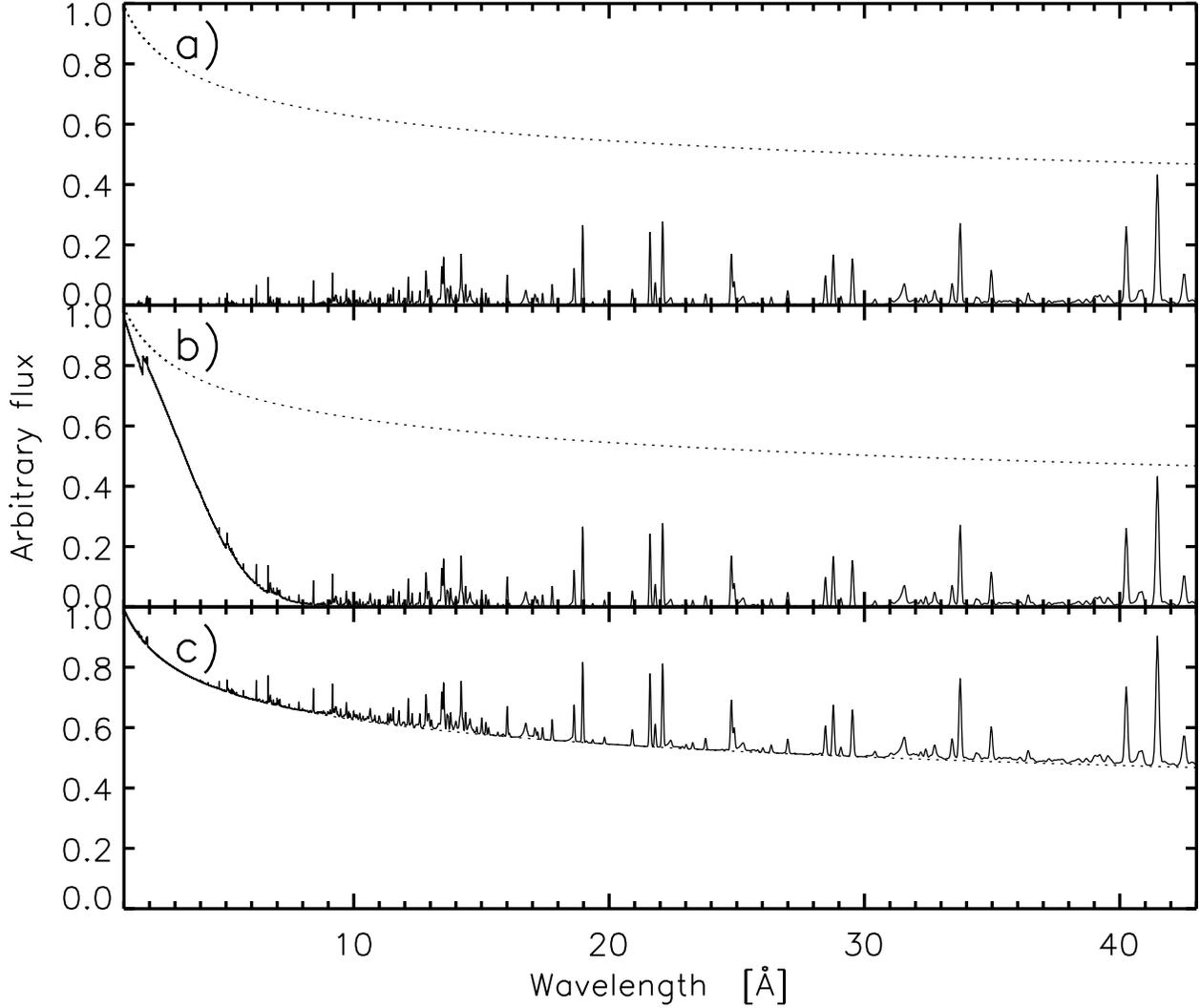,angle=90,width=8.4in}}
    \vspace{-0.45in}
\caption{``Reemission'' analogue (see Fig.~\ref{fig:cone}) to the
same absorption model used in the top panel of Fig.~\ref{fig:sy1}.  
For all spectra, we take 
$\sigma_v^{\mathrm{rad}}=\sigma_v^{\mathrm{trans}}=400$~km~s$^{-1}$, 
$\Gamma=1.8$, and $f=\Omega/4\pi=0.15$.  Here the three panels correspond to
differing amounts of obscuration along the observer's line-of-sight to the 
intrinsic continuum: a) Pure reemission spectrum (intrinsic continuum
is completely absorbed), b) Reemission spectrum plus moderately-absorbed 
intrinsic continuum (with neutral opacity towards intrinsic continuum 
equivalent to $N_{\mathrm{H}}=5.0\, $e22~cm$^{-2}$), and c) Reemission plus 
unabsorbed intrinsic continuum.
\label{fig:sy2}}
  \end{minipage}
\end{figure*}

Observations transverse to the cone yield line spectra resulting from
reemission in the cone.
We concentrate on the dominant reemission from H- and He-like ions 
for simplicity (the contributions due to lower-charge-state ions is 
straightforward to include).
The final spectrum of transitions to the ground state of all relevant 
K-shell ions is then simply the sum of all ionic contributions:
\begin{equation}
\mathrm{Reemission}(E)=\sum_{j,i} N_{z,j} A_{z,j}^{z,i} l_{j\rightarrow i}(E)+\Delta_{z-1}^{z}\frac{C_{z-1,g}^{z,g}}{C_{\mathrm{total}}}r(E),
\label{eq:reem}
\end{equation}
where $\Delta_{z-1}^{z}$ is the total recombination rate calculated from 
Eq.~\ref{eq:photrec} and $l_{j\rightarrow i}(E)$ is the appropriate Voigt 
line profile (normalized to 1) using velocity distribution 
$\sigma_v^{\mathrm{trans}}$ and $r(E)$ is the ground-state recombination 
profile (normalized to 1) convolved with $\sigma_v^{\mathrm{trans}}$.  In 
general, $\sigma_v^{\mathrm{trans}}$ may not necessarily be equal to 
$\sigma_v^{\mathrm{rad}}$.  Explicitly, and without velocity convolution, 
the recombination profile is simply:
\begin{equation}
r(E)\propto f(E-E_{\mathrm{th}};kT_{\mathrm{e}}) \frac{E^2}{\sqrt{E-E_{\mathrm{th}}}} \sigma_{\mathrm{PI}}(E),
\end{equation}
where $f(E-E_{\mathrm{th}};kT_{\mathrm{e}})$ is the Maxwellian distribution, 
evaluated at $E-E_{\mathrm{th}}$ and specified by temperature 
$kT_{\mathrm{e}}$, and $\sigma_{\mathrm{PI}}(E)$ is the photoionization cross 
section, with $\sigma_{\mathrm{PI}}(E)=0$ for $E<E_{\mathrm{th}}$.  The 
contribution from radiative and dielectronic 
recombination forming lower-ionization-state ions is also straightforward to 
include.  Examples of reemission spectra for differing amounts of obscuration 
towards the intrinsic continuum are shown in Fig.~\ref{fig:sy2}).

If the emission regions are moderately optically thick to their emission, then
the reemitted spectra will be modified.  Higher-order transitions 
($n\rightarrow1$ with $n>2$) will be degraded through multiple scattering to 
lower-order transitions (at sufficiently high optical depths, all higher order 
lines are converted to $n=2\rightarrow1$ transitions).  This
reprocessing can provide an estimate of the overall optical thickness 
of the medium.  In Fig.~13 of Kinkhabwala \etal\ (2002), the diagnostic value 
of this effect in H-like and He-like line series is illustrated.  The lines 
shown in that plot give the column density and velocity width in (assuming a 
roughly spherical medium) corresponding to 10\% conversion of the 
Ly$\beta$ photons to Ly$\alpha$ and, similarly, 10\% conversion of He$\beta$ 
to He$\alpha$ transitions for C through Mg (degradation percentage 
can be scaled arbitrarily in these plots).

\subsection{Absorption Plus Reemission}\label{sec:absreem}

Even if the ionization cone is viewed in absorption, there is some amount of 
the reprocessed emission which will be reemitted in the observer's direction 
(see Fig.~\ref{fig:cone}).  The total spectrum can be written as
$\mathrm{Spectrum}(E)= L(E) \mathrm{e}^{-\tau(E)}+\mathrm{Reemission}(E)$,
where $\mathrm{Reemission}(E)=\mathrm{Decay}(E)+\mathrm{Recombination}(E)$,
with $\mathrm{Decay}(E)$ giving the total contribution from radiative decay 
following photoexcitation and $\mathrm{Recombination}(E)$ giving the total 
contribution of both radiative and dielectronic recombination.
Both $\mathrm{Decay}(E)$ and $\mathrm{Reemission}(E)$ implicitly contain the 
details of absorption in the cone, as shown below.

In order to account for absorption of the reemitted spectrum in the cone
we must make some further assumptions.  First, we assume that the cone is 
sufficiently narrow so that the absorption factor pertaining to any region of 
the cone is just dependent on radius.  We define two such absorption factors:  
$\tau_o(E,r)=\int_{r_o}^{r} \sum_{\{Z,z\}} n_{Z,z,g}(r)\big[\sigma^R\big]_{Z,z,g}(E)$, 
representing absorption from the inner radius $r_o$ of the cone to $r$, and 
$\tau_f(E,r)=\int_r^{r_f} \sum_{\{Z,z\}} n_{Z,z,g}(r)\big[\sigma^R\big]_{Z,z,g}(E)$, 
representing absorption from $r$ to the outer radius of the cone $r_f$.  Clearly, the 
total opacity is just $\tau(E)=\tau_o(E,r)+\tau_f(E,r)$ for any value of $r$.

First, we derive a general expression for $\mathrm{Decay}(E)$.  
For simplicity, we assume that photoexcitation corresponds to coherent resonant 
scattering (in contrast to statements made in \S\ref{sec:master}).  
The relevant spectrum expressed as a simple integral over the cone can then be written
as:
\begin{eqnarray}
\mathrm{Decay}(E)&=&\sum_{\{Z,z\}}\int F(E,r) n_{Z,z,g}(r)\sum_j\big[\sigma^R\big]_{Z,z,g}^{Z,z,j}(E) \mathrm{e}^{-{\tau_f(r)}}r^2 dr d\Omega=\nonumber\\
&&\sum_{\{Z,z\}}\int \frac{L(E) \mathrm{e}^{-\tau_0(r,E)}}{4\pi r^2}  n_{Z,z,g}(r)\sum_j\big[\sigma^R\big]_{Z,z,g}^{Z,z,j}(E) \mathrm{e}^{-{\tau_f(r,E)}}r^2 dr d\Omega=\nonumber\\
&&\frac{\Omega}{4\pi}L(E) \mathrm{e}^{-\tau(E)}\sum_{\{Z,z\}} \sum_j\big[\sigma^R\big]_{Z,z,g}^{Z,z,j}(E)\int  n_{Z,z,g}(r)  dr=\nonumber\\
&&\frac{\Omega}{4\pi}L(E) \mathrm{e}^{-\tau(E)}\sum_{\{Z,z\}} \sum_j G_{Z,z}\big[\sigma^R\big]_{Z,z,g}^{Z,z,j}(E).
\end{eqnarray}

Accounting for absorption of the reemission contribution from recombination is
more difficult due to the need to integrate over each photoelectric edge to 
determine the photoionization (and, therefore, recombination) rate.  Here, we 
define 
$R_{Z,z}(E)$ to be the spectrum (properly normalized) of recombination 
(radiative and dielectronic) forming ion $Z,z$.  Total recombinations are 
balanced with photoionizations (resulting either from direct photoionization 
or indirect ionization via autoionization following 
photoexcitation; though only the contribution from the former is shown 
below for simplicity).  This yields:
\begin{eqnarray}
\mathrm{Recombination}(E)&=&\sum_{\{Z,z\}}R_{Z,z}(E)\int\bigg(\int F(E',r) n_{Z,z,g}(r)\sum_j\big[\sigma^R\big]_{Z,z,g}^{Z,z-1,j}(E')dE'\bigg) \mathrm{e}^{-{\tau_f(r)}}r^2 dr d\Omega=\nonumber\\
&&\frac{\Omega}{4\pi}\sum_{\{Z,z\}}R_{Z,z}(E)\int\bigg(\int L(E') \mathrm{e}^{-\tau_0(r,E')}\sum_j\big[\sigma^R\big]_{Z,z,g}^{Z,z-1,j}(E')dE'\bigg) \mathrm{e}^{-{\tau_f(r)}} n_{Z,z,g}(r) dr.
\label{eq:recomabs}
\end{eqnarray}
Here, the integral over energy in parentheses is still a function of $r$ through 
$\tau_o(r,E)$, requiring a costly double integral.  In order to speed up this 
calculation, we would like to split this double integral into two sequential 
integrations over $E$ and then $r$.  This requires approximating the opacity term
within the energy integral as some average opacity independent of radius, 
$\mathrm{e}^{-\tau_o(r,E)}\approx\mathrm{Opacity}(E)$.  
Employing once again the assumption that all ionic ratios are constant
throughout the cone, we obtain:
\begin{eqnarray}
\mathrm{Recombination}(E)=\frac{\Omega}{4\pi}\sum_{\{Z,z\}}R_{Z,z}(E)\bigg(\int L(E') \mathrm{Opacity}(E')\sum_j\big[\sigma^R\big]_{Z,z,g}^{Z,z-1,j}(E')dE'\bigg)\times\\
 \int \mathrm{e}^{-{\tau_f(r,E)}}      \frac{n_{Z,z,g}(r)}{\sum_{\{Z',z\}}{n_{Z',z,g}(r)\sigma^R_{Z',z}(E)}}\sum_{\{Z',z\}}n_{Z',z,g}(r)\sigma^R_{Z',z}(E)   dr 
\end{eqnarray}
Recalling that $d\tau_f(r,E)=-\sum_{\{Z',z\}}n_{Z',z,g}(r)\sigma^R_{Z',z}(E) dr$, we 
can integrate this expression (e.g., see 
Eqs.~\ref{eq:fractionalterm}--\ref{eq:Gthetatwo}), obtaining:
\begin{eqnarray}
\mathrm{Recombination}(E)=\frac{\Omega}{4\pi} \sum_{\{Z,z\}}G_{Z,z}R_{Z,z}(E)\frac{1-\mathrm{e}^{-\tau(E)}}{\tau(E)}\int L(E') \mathrm{Opacity}(E')\sum_j\big[\sigma^R\big]_{Z,z,g}^{Z,z-1,j}(E')dE'.
\label{eq:recomabs2}
\end{eqnarray}
Useful lower and upper limits to the recombination contribution are obtained 
by taking  $\mathrm{Opacity}(E)=\mathrm{e}^{-\tau(E)}$ and $\mathrm{Opacity}(E)=1$, 
respectively.  Spectra resulting from assumption of either limit can be 
calculated with \photoion\ (see Fig.~\ref{fig:sy1}).

\section{Discussion}\label{sec:dis}

The models presented in this paper have already received experimental 
verification through X-ray spectra of multiple accretion-powered
astrophysical sources.  In particular, application of our model to the 
remarkable \xmm\ spectrum of the brightest, prototypical Seyfert~2 galaxy, 
\ngc~1068 (Kinkhabwala \etal\ 2002), provided the first quantitative test of a photoionization 
code (e.g., the gratifying confirmation of the ratio of intercombination line 
to {\small RRC} in \ion{O}{7}).  In addition, the self-consistent inclusion of 
photoexcitation and resulting radiative decay allowed for discrimination 
between emission mechanisms in the spectrum of \ngc~1068, providing a robust 
upper limit to any additional contribution from hot plasmas.  
Observations of \ngc~1068 also provided the motivation for the seemingly 
{\em ad hoc} assumption taken in \S\ref{sec:cone} that all ionic ratios are 
constant throughout the cone (Brinkman \etal\ 2002; Ogle \etal\ 2003), as
was explained in \S\ref{sec:intro}.  We expect 
that \photoion\ (and associated codes {\small PHSI}, {\small MPABS}, and 
{\small SIABS}) will continue to be of use in the description of 
astrophysical photoionized plasmas existing in a variety of sources, including 
AGN, X-ray binaries, cataclysmic variables, stellar winds of early-type stars, 
GRB afterglows, and the IGM.

Lastly, as mentioned in \S\ref{sec:atom}, radiative and autoionizing decay 
rates (as well as the resulting fluorescence intensities) for
levels capable of autoionization have not yet been properly included.  
Especially important are the Fe K-shell 
transitions in Fe M-shell and L-shell ions.  The number of atomic levels 
involved in these calculations is 
large, but not prohibitive.  These represent the final atomic calculations 
necessary to ``complete'' \photoion, at least in terms of the assumptions 
taken in this paper.  These rates should eventually be included in a future 
version of the code.

\acknowledgments{AK acknowledges useful discussions with D.W. Savin. 
The Columbia University team is supported by \nasa.  AK acknowledges 
additional support from an \nsf\ Graduate Research Fellowship and \nasa\
\gsrp\ fellowship.  EB was supported by the Yigal-Alon Fellowship and 
by the GIF Foundation under grant \#2028-1093.7/2001.
MS and MFG were partially supported by {\small NASA} through {\it   Chandra} 
Postdoctoral Fellowship Award Numbers {\small PF}01-20016 and 
{\small PF}01-10014, respectively, issued by the {\it Chandra} X-ray 
Observatory Center, which is operated by the Smithsonian Astrophysical 
Observatory for and behalf of {\small NASA} under contract 
{\small NAS}8-39073.}

\clearpage

\bibliographystyle{apj}

\end{document}